\documentclass{aa}  

\usepackage{graphicx}
\usepackage{txfonts}

\usepackage[colorlinks=true, allcolors=blue]{hyperref}
\makeatletter
\renewcommand*\aa@pageof{, page \thepage{} of \pageref*{LastPage}}
\makeatother

\newcommand{\bfk}{\boldsymbol{k}}
\newcommand{\bfd}{\boldsymbol{d}}
\newcommand{\bfS}{\boldsymbol{S}}
\newcommand{\bfp}{\boldsymbol{p}}

\begin{document}

   \title{X-ray polarisation in AGN circumnuclear media}

   \subtitle{Polarisation framework and 2D torus models}

   \titlerunning{X-ray polarisation in AGN circumnuclear media}

   \authorrunning{B.\ Vander Meulen et al.}

   \author{Bert Vander Meulen
          \inst{1}
          \and
          Peter Camps
          \inst{1}
          \and
          {\DJ}or{\dj}e {Savi{\'c}}
          \inst{2, 3}
          \and
          Maarten Baes
          \inst{1}
          \and
          Giorgio Matt
          \inst{4}
          \and
          Marko Stalevski
          \inst{1, 3}
          }
   \institute{Sterrenkundig Observatorium, Universiteit Gent, Krijgslaan 281 S9, 9000 Gent, Belgium\\
              \email{bert.vandermeulen@ugent.be}
              \and
               Institut d’Astrophysique et de Géophysique, Université de Liège, Allée du 6 Août 19c, 4000 Liège, Belgium
              \and
              Astronomical Observatory, Volgina 7, 11060 Belgrade, Serbia
              \and
              Dipartimento di Matematica e Fisica, Universitá degli Studi Roma Tre, via della Vasca Navale 84, 00146 Roma, Italy
        }
   \date{Received May 17, 2024; accepted July 29, 2024}


  \abstract
   {Cold gas and dust reprocess the central X-ray emission of active galactic nuclei (AGN), producing characteristic spectro-polarimetric features in the X-ray band. The recent launch of IXPE allows for observations of this X-ray polarisation signal, which encodes unique information on the parsec-scale circumnuclear medium of obscured AGN. However, the models for interpreting these polarimetric data are under-explored and do not reach the same level of sophistication as the corresponding spectral models.}
   {We aim at closing the gap between the spectral and spectro-polarimetric modelling of AGN circumnuclear media in the X-ray band by providing the tools for simulating X-ray polarisation in complex geometries of cold gas alongside X-ray spectra.}
   {We lay out the framework for X-ray polarisation in 3D radiative transfer simulations and provide an implementation to the 3D radiative transfer code SKIRT, focussing on (de)polarisation due to scattering and fluorescent re-emission. As an application, we explored the spectro-polarimetric properties of a 2D toroidal reprocessor of cold gas, modelling the circumnuclear medium of AGN.}
   {For the 2D torus model, we find a complex behaviour of the polarisation angle with photon energy, which we interpret as a balance between the reprocessed photon flux originating from different sky regions, with a direct link to the torus geometry. We calculated a large grid of AGN torus models and demonstrated how spatially resolved X-ray polarisation maps could form a useful tool for interpreting the geometrical information that is encoded in IXPE observations. With this work, we release high-resolution AGN torus templates that simultaneously describe X-ray spectra and spectro-polarimetry for observational data fitting with XSPEC.}
   {The SKIRT code can now model X-ray polarisation simultaneously with X-ray spectra and provide synthetic spectro-polarimetric observations for complex 3D circumnuclear media, with all features of the established SKIRT framework available.}


   \maketitle
%

\section{Introduction}
\label{sec:intro}
Active galactic nuclei (AGN) are the compact central regions of massive galaxies whose excessive brightness across the electromagnetic spectrum is powered by the accretion of gas and dust onto a supermassive black hole (SMBH) \citep{lyndenbell69, rees84}. AGN are the most luminous persistent sources in the Universe and play a crucial role in galaxy evolution by pumping energy and momentum into the interstellar gas and launching powerful jets and outflows. AGN feedback is one of the most important and least understood ingredients of galaxy evolution theories, with several observed correlations indicating that AGN and their host galaxy co-evolve and regulate each other’s growth \citep{magorrian98, ferrarese00, gebhardt00, marconi03}.

According to the unified AGN structure model, the observed dichotomy in AGN types is explained by a large-scale toroidal structure of gas and dust in the equatorial plane, which causes line-of-sight obscuration, depending on the observer's viewing angle \citep{antonucci93, urry95, netzer15}. This `torus' of gas and dust is then responsible for the extinction at optical and UV wavelengths, the thermal dust re-emission in the infrared, and reprocessing in the X-ray band. Furthermore, it explains the absence of broad optical lines in obscured AGN and the appearance of these lines in polarised light \citep{antonucci85}. This circumnuclear torus medium could further be important as an accretion reservoir fuelling the active SMBH or as a direct probe on AGN feedback \citep{ramosalmeida17}.

Lately, a new picture has emerged of the dust structure in local active galaxies, challenging the classical `dusty torus' paradigm. Spectral modelling suggests that the circumnuclear medium has a more complex three-dimensional structure with clumps and filaments \citep{ramosalmeida09, honig10, stalevski12, stalevski16, balokovic14, buchner19}, which is further supported by observations of AGN variability \citep{risaliti02, markowitz14, buchner19, ricci22, torresalba23}. Furthermore, high-angular resolution imaging in the mid-infrared has shown that the circumnuclear medium is also extended in the polar direction, as opposed to a purely equatorial torus  \citep{tristram07, tristram14, honig12, honig13, burtscher13, lopezgonzaga14, lopezgonzaga16, asmus16, stalevski17, stalevski19, leftley18, asmus19, isbell22, isbell23, leist24, haidar24}.

The AGN circumnuclear medium could be further explored through X-ray observations, as most AGN spectra show a strong X-ray component, which is produced by Compton up-scattering in a corona of hot electrons close to the SMBH \citep[][]{haardt91}. These coronal X-rays are then reprocessed by the circumnuclear material, producing characteristic spectral features that form a powerful probe regarding the distribution of gas and dust in local AGN \citep{ricci14, asmus15, ichigawa19}. X-rays have a high penetrating power, and therefore they shed light on the most obscured episodes of SMBH accretion. Indeed, a large population of obscured AGN is only revealed through X-ray observations, with spectra that are shaped by reprocessing in the AGN torus \citep{ricci17}. As many of these sources cannot be spatially resolved, they contribute to the cosmic X-ray background \citep{gilli07}.

The most prominent features of X-ray reprocessing by the AGN torus are the narrow Fe K$\alpha$ line at $6.4~\text{keV}$ \citep[][]{mushotzky78, nandra89, pounds89, fukazawa11} and the Compton reflection hump peaking at about $30~\text{keV}$ \citep[][]{matt91}. These features directly probe the cold gas and dust surrounding AGN and are commonly used to constrain the geometry of the reprocessing medium \citep[][]{gupta21, yamada21, osorioclavijo22, torresalba23}. Indeed, recent radiative transfer studies have demonstrated that X-ray spectra of obscured AGN carry detailed information on the 3D distribution of obscuring gas and dust, tracing clumpy structures \citep{buchner19, vandermeulen23} and polar extended material \citep{mcKaig22} in the circumnuclear medium, even outside of the line of sight.

Two additional observables can be extracted from the X-ray emission emerging from obscured AGN when dedicated polarisation instrumentation is available. The polarisation angle and polarisation degree encode complementary information on the circumnuclear medium, which can be used to constrain the geometry of the torus and its orientation relative to the host galaxy. The recent launch of the Imaging X-ray Polarimetry Explorer \citep[IXPE;][]{weisskopf22} introduced X-ray polarimetry as a new tool to study AGN in the $2-8~\text{keV}$ band, with five radio-quiet AGN that have been observed in the last two years: MCG-05-23-16 \citep{marinucci22, serafinelli23, tagliacozzo23}, the Circinus galaxy \citep{ursini23}, NGC4151 \citep{gianolli23}, IC4329A \citep{ingram23}, and NGC1068 \citep{marin24}. The next generation of X-ray polarisation missions, with the X-ray Polarimeter Satellite (XPoSat) \citep[$8-30~\text{keV}$,][]{ghosh23} and the enhanced X-ray Timing and Polarization mission (eXTP) \citep[$2-8~\text{keV}$,][]{zhang16, zhang19}, further indicate a promising future for X-ray polarimetry.

The recent developments in observational X-ray polarimetry motivate the need for more advanced polarisation models that are based on 3D radiative transfer simulations. However, spectro-polarimetric X-ray models for the AGN torus are under-explored and do not reach the same level of sophistication as the corresponding X-ray spectral models \citep[e.g.][]{murphy09, ikeda09, brightman11, odaka11, odaka16, liu14, furui16, paltani17, balokovic18, balokovic19, tanimoto19, buchner19, buchner21, ricci23, vandermeulen23}.

Historically, one of the first X-ray polarisation models was the wedge torus model by \citet{ghisellini94}, which has been used for the interpretation of the $772~\text{ks}$ of IXPE data on Circinus AGN \citep{ursini23} and for the exploration of the binary geometry of GRS 1915+105 \citep{ratheesh21}. However, this model allows for little geometrical flexibility.

The STOKES code \citep{goosmann07, marin12, marin15, rojaslobos18, marin18b} on the other hand offers a more flexible radiative transfer framework, and it has been used to make polarisation predictions for a range of 3D torus geometries \citep{goosmann11, marin13, marin16, marin17}. STOKES applies a Monte Carlo technique to model scattering-induced polarisation in the X-ray, UV, optical, and infrared bands. Recently, the STOKES code has been used to model X-ray polarisation in a parsec-scale equatorial torus, assuming neutral and partially ionised gas \citep{marin18c,marin18a,podgorny22,podgorny23a,podgorny23b,podgorny24c, podgorny24b, podgorny24a}, which has been applied to the $1.15~\text{Ms}$ IXPE observation of NGC1068 \citep{marin24}.

Most recently, the MONACO code \citep{odaka11, odaka16} has been used to predict the X-ray polarisation signal of Circinus AGN, which was compared to the IXPE observational data \citep{tanimoto23}. MONACO, which builds on the Geant4 simulation toolkit \citep{agostinelli03, allison06, allison16}, implements X-ray polarisation in cold neutral material, which has been applied to post-process the 3D hydrodynamical torus simulations by \citet[][]{wada16} modelling the circumnuclear gas in the Circinus galaxy.

With this work, we aim to close the gap between the X-ray spectral modelling and X-ray spectro-polarimetric modelling of AGN circumnuclear media by setting up 3D radiative transfer simulations that simultaneously predict X-ray polarisation and X-ray spectra. For this, we focus on the well-established SKIRT code \citep{camps20} so that our simulations are highly efficient in terms of computational runtime, allowing complex 3D models to be explored in a short time. Furthermore, the SKIRT code offers an unmatched geometrical flexibility for setting up radiative transfer simulations in full 3D, which has now become available to X-ray polarisation modelling. This work introduces the necessary tools for modelling X-ray polarisation in a general 3D context and presents an implementation to the SKIRT code, which is publicly available. As a first application, we explored the spectro-polarimetric properties of a simple 2D torus model. Future work will focus on more complex 3D geometries beyond the classical AGN torus.

The goal of this work is to lay out the X-ray polarisation framework for 3D radiative transfer codes and investigate the X-ray polarisation observables of a classical AGN torus model. The outline for this paper is as follows: In Sect.~\ref{sec:framework}, we introduce the polarisation framework. In Sect.~\ref{sec:implementation}, we provide an implementation to the 3D radiative transfer code SKIRT. In Sect.~\ref{sec:torus}, we study the polarisation observables of a 2D torus model and present the corresponding torus templates for observational data fitting. We discuss our findings in Sect.~\ref{sec:discussion} and summarise our results in Sect.~\ref{sec:summary}.

\section{X-ray polarisation framework}
\label{sec:framework}
\subsection{Stokes vector formalism}
\label{sec:stokes}
In Monte Carlo radiative transfer simulations, the polarisation state of the (discretised) radiation field is most conveniently characterised in terms of the four Stokes parameters \citep{stokes51}. Together, these parameters form a Stokes vector $\bfS$, which is traced and updated throughout the simulated transfer medium for each `photon packet':
\begin{equation}
    \bfS = \begin{pmatrix} I \\ Q \\ U \\ V \end{pmatrix}.
\end{equation}
Stokes parameter $I$ represents the total intensity of the photon packet, while $Q$ and $U$ describe linear polarisation, and $V$ describes circular polarisation. The four Stokes parameters encode all polarisation information of the radiation field (except for its phase), so that the linear polarisation degree $P_{\text{L}}$ can be reconstructed as
\begin{equation}
    P_{\text{L}} = \frac{\sqrt{Q^2 + U^2}}{I}, \label{eq:PL}
\end{equation}
and the linear polarisation angle $\gamma$ can be found as
\begin{equation}
    \gamma = \frac12\arctan_2\left(\frac{U}{Q}\right), \label{eq:gamma}
\end{equation}
with $\arctan_2$ being the inverse tangent function that preserves the quadrant, also noted as \textsc{atan2(U, Q)}.

Stokes vectors $\bfS$ are defined relative to a reference direction $\bfd$ that is perpendicular to the photon propagation direction $\bfk$. Positive $Q$-values describe linear polarisation in the reference direction, while negative $Q$-values describe linear polarisation perpendicular to this reference direction. Equivalently, Stokes $U$ describes linear polarisation along orthogonal axes rotated over  $45^\circ$ with respect to the reference direction. One has
\begin{subequations}\label{eq:QUinterpretation}
\begin{align}
Q &= I P_{\text{L}} \cos2\gamma,\\
U &= I P_{\text{L}} \sin2\gamma.
\end{align}
\end{subequations}
For the remainder of this work, we focus on linear polarisation, as IXPE observations do not capture the Stokes $V$-component. Nevertheless, our polarisation framework includes a functional implementation of circular polarisation (which is not discussed).

When the reference direction $\bfd$ is rotated by an angle $\varphi$ about $\bfk$ to a new reference direction $\bfd_\text{rot}$, the Stokes parameters transform as
\begin{equation}
\label{eq:applyRotation}
\bfS_\text{rot} = {\textbf{R}}(\varphi)\,\bfS,
\end{equation}
with ${\textbf{R}}(\varphi)$ being the rotation matrix, which is given as
\begin{equation}
{\textbf{R}}(\varphi) =
\begin{pmatrix}
1 & 0 & 0 & 0 \\
0 & \cos2\varphi & \sin2\varphi & 0 \\
0 & -\sin2\varphi & \cos2\varphi & 0 \\
0 & 0 & 0 & 1
\end{pmatrix},
\label{eq:rotation}
\end{equation}
mixing the $Q$ and $U$ parameters. Combining Eq.~(\ref{eq:QUinterpretation}), (\ref{eq:applyRotation}) and (\ref{eq:rotation}), we obtain
\begin{subequations}\label{eq:QUrot}
\begin{align}
Q_\text{rot} &= I P_{\text{L}} \cos2\left(\gamma - \varphi\right),\\
U_\text{rot} &= I P_{\text{L}} \sin2\left(\gamma - \varphi\right),
\end{align}
\end{subequations}
so that we have ${P_{\text{L}}}_\text{rot}=P_{\text{L}}$ and $\gamma_\text{rot}=\gamma - \varphi$. This rotation transformation is important to describe scattering interactions relative to a reference direction that lays in the scattering plane and to record Stokes vectors relative to the north direction of the observer frame (following the IAU conventions).

\begin{figure}
    \centering
	\includegraphics[width=0.8\columnwidth, trim={5cm 7cm 9cm 13cm},clip]{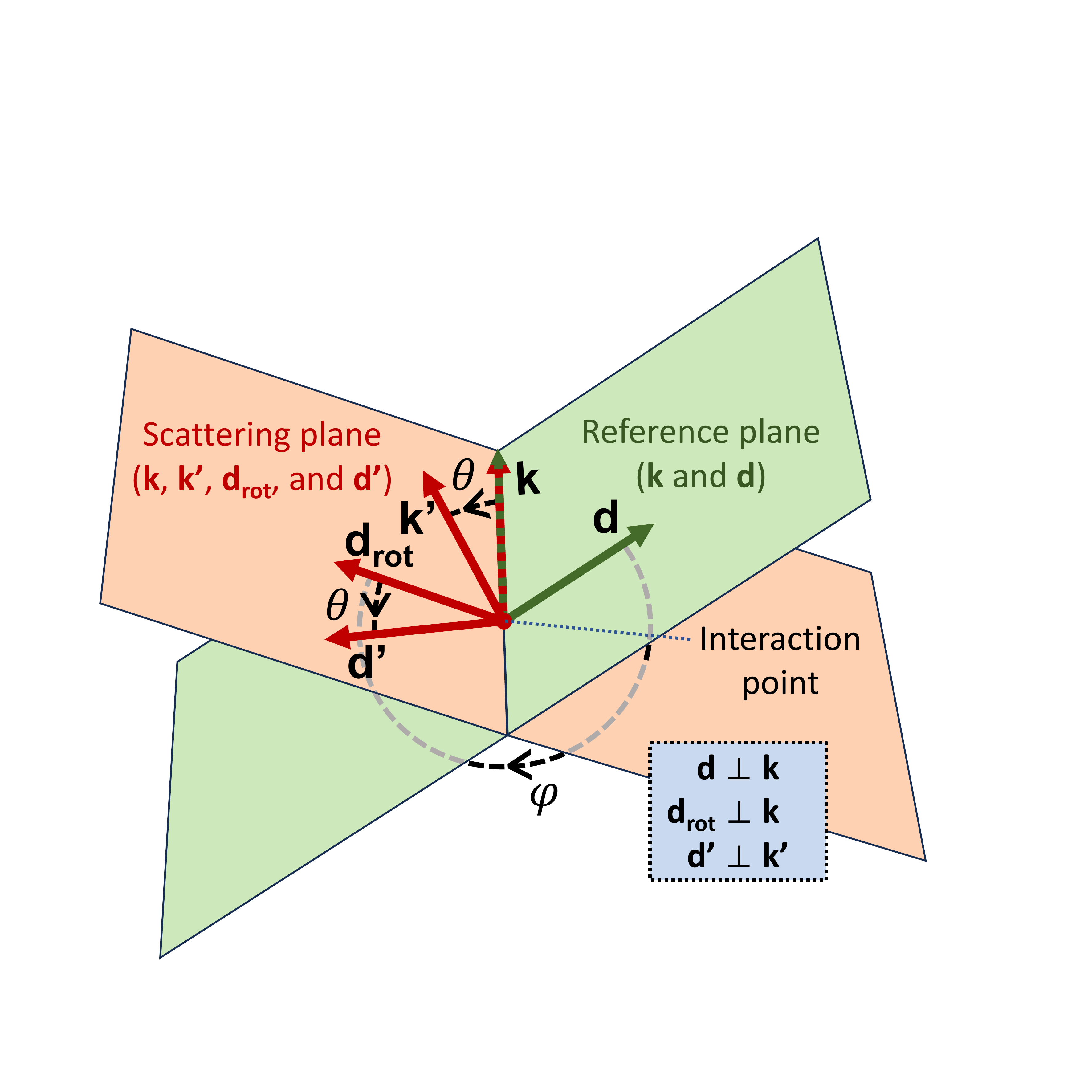}
    \caption{Three-dimensional scattering geometry visualising the unit direction vectors, unit reference vectors, and rotation angles. The reference plane is defined by the incoming photon propagation direction $\bfk$ and the initial polarisation reference direction $\bfd$. The scattering plane is defined by the incoming photon direction $\bfk$ and the outgoing photon direction $\bfk'$, and also contains $\bfd_\text{rot}$ and $\bfd'$ (see text for more details).}
    \label{fig:scatteringgeometry}
\end{figure}
In polarised radiative transfer, photon interactions generally depend on the polarisation state, meaning that the processes are described in terms of the Stokes parameters of that photon. In addition, scattering can modify the polarisation state of a photon, which can be described as a matrix multiplication on $\bfS$ with the corresponding M\"uller scattering matrix ${\textbf{M}}$. In this work, M\"uller matrices ${\textbf{M}}(\theta, x)$ are a function of the scattering angle $\theta$ and the incoming photon energy $x = E/m_\text{e}c^2$. Considering that the reference direction of the incoming photon should first be rotated to an intermediate reference direction $\bfd_\text{rot}$ in the scattering plane, the Stokes vector $\bfS'$ after scattering can be obtained as
\begin{equation}\label{eq:genScatter}
    \bfS'={\textbf{M}}(\theta,x)\ \textbf{R}(\varphi)\ \bfS,
\end{equation}
where $\varphi$ is the angle by which the initial reference direction $\bfd$ must be rotated about $\bfk$ to end up in the scattering plane as $\bfd_\text{rot}$ (see Fig.~\ref{fig:scatteringgeometry}). Finally, we note that Stokes vector $\bfS'$ refers to a new reference direction $\bfd'$, which is just the intermediate reference direction $\bfd_\text{rot}$, rotated over the scattering angle $\theta$ in the scattering plane (i.e.\ a rotation about the scattering plane normal $\bfk\times\bfd_\text{rot}$), to assure that the new reference direction $\bfd'$ is perpendicular to the new propagation direction $\bfk'$. The 3D geometry of a scattering interaction is shown in Fig.~\ref{fig:scatteringgeometry}, with unit direction vectors, unit reference vectors, and rotation angles indicated. For more details, we refer to \citet[][]{peest17}. Applying Rodrigues' rotation formula, we obtain an expression for $\bfd'$ as
\begin{equation}\label{eq:dprime}
    \bfd'=\cos\varphi\cos\theta\:\bfd + \sin\varphi\cos\theta  \left(\bfk \times \bfd\right) -\sin\theta \: \bfk.
\end{equation}

Various normalisation conventions are in use for the M\"uller matrices ${\textbf{M}}$ (e.g.\ different conventions can be found in \citet{depaola03}, \citet{kasen06} and \citet{peest17}). However, polarisation observables such as $P_{\text{L}}$ and $\gamma$ do not depend on this absolute normalisation factor, as they are calculated from Stokes parameter ratios, causing the normalisation factor to cancel out. The absolute polarised fluxes $Q$ and $U$ can be recovered from the normalised total flux $I$ (similar to Eq.~(\ref{eq:QUinterpretation})), which is discussed in Sect.~\ref{sec:compton}.

\subsection{X-ray radiative processes}
\label{sec:processes}
In this work, we focus on radiative transfer simulations in cold atomic gas, modelling the circumnuclear medium that is causing most of the X-ray extinction in obscured AGN. Furthermore, this material is responsible for the distinct X-ray reflection features observed in Compton-thick AGN, such as the narrow Fe K$\alpha$ line at $6.4~\text{keV}$ and the Compton reflection hump at about $30~\text{keV}$. In particular, we consider photo-absorption by neutral atomic gas with self-consistent fluorescent line emission and bound-electron scattering.

Photo-absorption does not depend on the polarisation state of the incoming photon. Therefore, it can be implemented using the standard \citet{verner95, verner96} cross-sections and a custom abundance table. Subsequently, the absorbed photon energy can be re-emitted as a fluorescent line photon, with a probability given by the fluorescent yield of that atom \citep[see e.g.][]{perkins91}. These line photons are unpolarised, regardless of the initial polarisation state of the absorbed photon \citep{compton28}. Fluorescence thus resets the Stokes vector, and the sequence of photo-absorption followed by fluorescent re-emission effectively acts as a depolarisation process.

In cold atomic gas, X-ray photons are scattered by the electrons that are bound to the neutral gas atoms, which can be described in terms of atomic form factors and incoherent scattering functions \citep[see e.g.][]{hubbell75}. However, this bound-electron scattering can be reasonably well approximated as Compton scattering on $Z$ free electrons per neutral gas atom, which we assume in this work. For \citet{anders89} solar gas abundances, this corresponds to $1.21$ free electrons per H-atom, which changes for other abundance tables. This free-electron approximation, applied to X-ray polarisation in a general 3D context, is the focus of this work, and we refer to future work for a treatment of polarised bound-electron scattering \citep[following the prescriptions of][]{matt96}.

In the next section, we describe the details of polarised Compton scattering (approximating bound-electron scattering), which is polarisation dependent and updates the polarisation state of the incoming photon. In particular, we focus on the total scattering cross-section and the scattering phase function for polarised photons and the M\"uller matrix for updating the Stokes parameters after a scattering interaction. This framework could also be used to model media of free electrons only, in addition to the cold gas that is the focus here.

\subsection{Polarised Compton scattering}
\label{sec:compton}
Compton scattering describes the inelastic scattering of high-energy photons on free electrons, which is often used to approximate bound-electron scattering in cold atomic gas \citep[see][for a discussion]{vandermeulen23}. The M\"uller matrix for Compton scattering is given by \citet{fano49}, using the same sign convention as the IAU \citep[see][]{peest17}:
\begin{equation}\label{eq:matrix}
{\textbf{M}}(\theta, x) \propto
\begin{pmatrix}
{\mathrm{S}}_{11}(\theta, x) & {\mathrm{S}}_{12}(\theta, x) & 0 & 0 \\
{\mathrm{S}}_{12}(\theta, x) & {\mathrm{S}}_{22}(\theta, x) & 0 & 0 \\
0 & 0 &  {\mathrm{S}}_{33}(\theta, x) & 0 \\
0 & 0 & 0 & {\mathrm{S}}_{44}(\theta, x)
\end{pmatrix},
\end{equation}
with $\theta$ being the scattering angle and $x = E/m_ec^2$ the photon energy scaled to the electron rest energy. Moreover, we assume an isotropic distribution for the electron spin directions, which causes the non-diagonal matrix elements in the fourth row and fourth column to be zero \citep{fano49}. The non-zero matrix elements of ${\textbf{M}}(\theta, x)$ are
\begin{subequations}
\label{eq:matrixelements}
\begin{align}
    {\mathrm{S}}_{11}(\theta, x) &= C^3(\theta, x) + C(\theta, x) -C^2(\theta, x)\sin^2\theta,\\
{\mathrm{S}}_{12}(\theta, x) &= -C^2(\theta, x)\sin^2\theta,\\
{\mathrm{S}}_{22}(\theta, x) &= C^2(\theta, x) \left(1+\cos^2\theta\right),\\
{\mathrm{S}}_{33}(\theta, x) &= 2C^2(\theta, x)\cos\theta,\\
{\mathrm{S}}_{44}(\theta, x) &= \left(C^3(\theta, x) + C(\theta, x)\right) \cos\theta,
\end{align}
\end{subequations}
with $C(\theta, x)$ being the Compton factor:
\begin{equation}
    C(\theta, x) = {\Big({1+x \, \left(1-\cos \theta\right)}\Big)}^{-1}. \label{comptonfactor}
\end{equation}
In the low energy limit, $C(\theta, x)\to 1$, so that Eq.~(\ref{eq:matrix}) converges to the M\"uller matrix for non-relativistic Thomson scattering.

Matrix element ${\mathrm{S}}_{11}(\theta, x)$ relates the scattered intensity $I'$ to the incoming intensity $I$ in case of unpolarised photons and is proportional to the \citet{klein29} differential cross-section for unpolarised Compton scattering. For (partially) polarised radiation, ${\mathrm{S}}_{12}(\theta, x)$ introduces an azimuthal modulation to the Klein-Nishina formula, which depends explicitly on the polarisation state of the incoming photon. Together, ${\mathrm{S}}_{11}(\theta, x)$ and ${\mathrm{S}}_{12}(\theta, x)$ make up the differential cross-section for polarised Compton scattering:
\begin{equation}
    \frac{d\sigma}{d\Omega}(\theta, \varphi, x, \bfS) = \frac{3\sigma_\text{T}}{16\pi} \left[{\mathrm{S}}_{11}(\theta, x) + {\mathrm{S}}_{12}(\theta, x)\,P_\text{L}\cos2(\varphi-\gamma)\right], \label{eq:dcs}
\end{equation}
with $\sigma_\text{T} \approx 6.65 \times 10^{-25}~\text{cm}^2$ being the Thomson cross-section and $\varphi$ the azimuthal scattering angle relative to the polarisation reference direction $\bfd$ (so that $\varphi-\gamma$ is the azimuthal scattering angle relative to the incoming polarisation direction $\bfp$).

For fully polarised photons ($P_\text{L}=1$), Eq.~(\ref{eq:dcs}) reduces to
\begin{equation}
    \frac{d\sigma}{d\Omega}(\theta, \varphi, x, \bfS)\bigg\rvert_{P_\text{L}=1} = \frac{3 \sigma_\text{T}}{16\pi} \left[ C^3 + C -2 C^2\sin^2\theta \cos^2(\varphi-\gamma)\right], \label{eq:lin_dcs}
\end{equation}
with $C \equiv C(\theta, x)$. In the IXPE range, the angular dependence of $C(\theta, x)$ is weak, and Eq.~(\ref{eq:lin_dcs}) is roughly symmetric around the incoming polarisation direction $\bfp$, as the factor $\sin\theta\, \cos(\varphi-\gamma)$ is just the cosine between $\bfk'$ and $\bfp$. For a full derivation from quantum electrodynamics, we refer to \citet{heitler54}.

By integrating Eq.~(\ref{eq:dcs}) over the unit sphere, we obtain the total cross-section for polarised Compton scattering as
\begin{equation}
    \sigma(x) = \frac{3\, \sigma_\text{T}}{4} \left[\frac{1+x}{(1+2x)^2} + \frac{2}{x^2} + \left(\frac{1}{2x}-\frac{x+1}{x^3}\right) \ln(2x+1) \right],
    \label{eq:totalcompton}
\end{equation}
which is the exact same expression as the total cross-section for unpolarised Compton scattering. This cross-section does not depend on the polarisation state, meaning that polarisation effects do not influence the Compton scattering efficiency. Furthermore, this cross-section is roughly constant over the IXPE energy range, decreasing from $0.99\,\sigma_\text{T}$ at $2~\text{keV}$ to $0.97\,\sigma_\text{T}$ at $8~\text{keV}$.

\begin{figure*}
    \centering
	\includegraphics[width=\hsize]{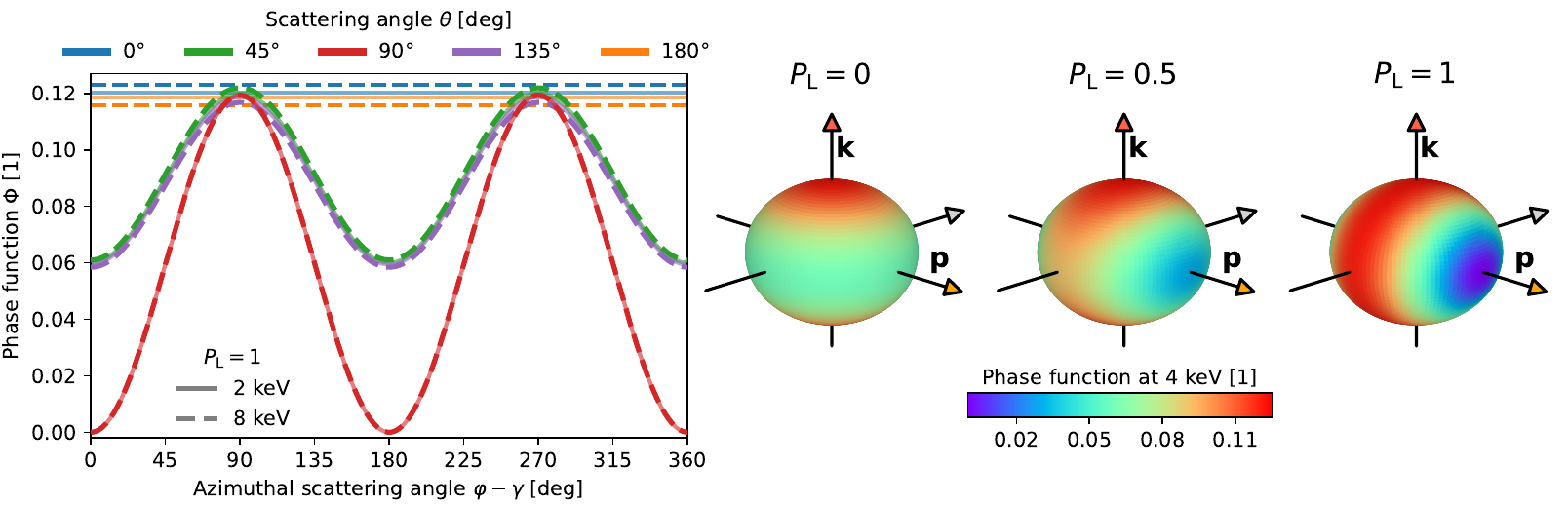}
    \caption{Compton scattering phase function (left) Phase function for fully polarised photons ($P_\text{L}=1$), with $\varphi-\gamma$ being the azimuthal scattering angle relative to the incoming polarisation direction. The solid and dashed lines correspond to energies of $2$ and $8~\text{keV}$, respectively. (right) For unpolarised photons ($P_\text{L}=0$), the Compton phase function is symmetric around the incoming direction $\bfk$, while for fully polarised photons ($P_\text{L}=1$), it is symmetric around the incoming polarisation direction $\bfp$. For partially polarised photons, the phase function is truly asymmetric. }
    \label{fig:PhaseFunction}
\end{figure*}
Normalising Eq.~(\ref{eq:dcs}) on the unit sphere, we obtain the phase function for polarised Compton scattering, shown in Fig.~\ref{fig:PhaseFunction}:
\begin{align}
        \Phi(\theta, \varphi, x, \bfS) &= \frac{3\sigma_\text{T}}{16\pi\sigma(x)} \left[{\mathrm{S}}_{11}(\theta, x) + {\mathrm{S}}_{12}(\theta, x)\,P_\text{L}\cos2(\varphi-\gamma)\right]\notag \\
                                       &= \Psi(\theta, x) - \frac{3\sigma_\text{T}\, C^2(\theta, x)}{16\pi\sigma(x)}\sin^2{\theta}\,P_\text{L} \cos2(\varphi-\gamma).\label{eq:Comptonphasefunction}
\end{align}
This phase function $\Phi(\theta, \varphi, x, \bfS)$ is the sum of the standard phase function for unpolarised Compton scattering $\Psi(\theta, x)\propto {\mathrm{S}}_{11}(\theta, x)$ and an azimuthal modulation that is characteristic for polarised Compton scattering. The azimuthal cosine modulation introduces a bias towards scattering into the plane perpendicular to the polarisation direction ($\varphi= \gamma + 90^\circ$ and $\varphi= \gamma + 270^\circ$), with an amplitude that is roughly proportional to $\sin^2{\theta} \times P_\text{L}$. Therefore, the modulation is most pronounced at $\theta=90^\circ$, where it maximally reduces the probability for scattering into the polarisation direction ($\varphi= \gamma$ or $\varphi= \gamma + 180^\circ$). As expected, the modulation strength is linearly proportional to the polarisation degree $P_\text{L}$, as only polarised photons experience the modulation effect.

The left panel of Fig.~\ref{fig:PhaseFunction} shows the Compton phase function at $2~\text{keV}$ (solid line) and $8~\text{keV}$ (dashed line), illustrating how the azimuthal modulation does not depend on the photon energy in the IXPE range. The only difference is the slightly stronger preference for forward scattering at higher energies, which is regulated by the polarisation-independent phase function term $\Psi(\theta, x)$, as is the case for unpolarised Compton scattering. The right panel of Fig.~\ref{fig:PhaseFunction} visualises the scattering phase function for three (incoming) polarisation states, illustrating how the phase function for unpolarised photons ($P_\text{L}=0$) is symmetric around the incoming photon direction $\bfk$, while for fully polarised photons ($P_\text{L}=1$), the phase function is almost perfectly symmetric around the incoming polarisation direction $\bfp$. For partially polarised photons, the scattering phase function is truly asymmetric, as illustrated for $P_\text{L}=0.5$ on the right panel of Fig.~\ref{fig:PhaseFunction}.

After a Compton scattering interaction, the Stokes parameters of the scattered photon are updated as
\begin{equation}
\label{eq:updateStokes}
    \begin{pmatrix} I' \\ Q' \\ U' \\ V' \end{pmatrix}
    \propto
    \begin{pmatrix} {\mathrm{S}}_{11}(\theta, x) \, I + {\mathrm{S}}_{12}(\theta, x) \, \left(\cos2\varphi\,Q+\sin2\varphi\,U\right)\\ {\mathrm{S}}_{12}(\theta, x) \, I + {\mathrm{S}}_{22}(\theta, x)\, \left(\cos2\varphi\,Q+\sin2\varphi\,U\right) \\ {\mathrm{S}}_{33}(\theta, x)\, \left(-\sin2\varphi\,Q+\cos2\varphi\,U\right) \\ {\mathrm{S}}_{44}(\theta, x)\,V \end{pmatrix},
\end{equation}
combining Eq.~(\ref{eq:rotation}), (\ref{eq:genScatter}), and (\ref{eq:matrix}). In addition, the photon energy is reduced to $E'=C(\theta, x) \times E$, and the updated Stokes vector $\bfS'$ is normalised so that $I'=C(\theta, x) \times I$, to comply with the conservation of four-momentum for the photon-electron pair.

\begin{figure}
    \centering
	\includegraphics[width=\columnwidth]{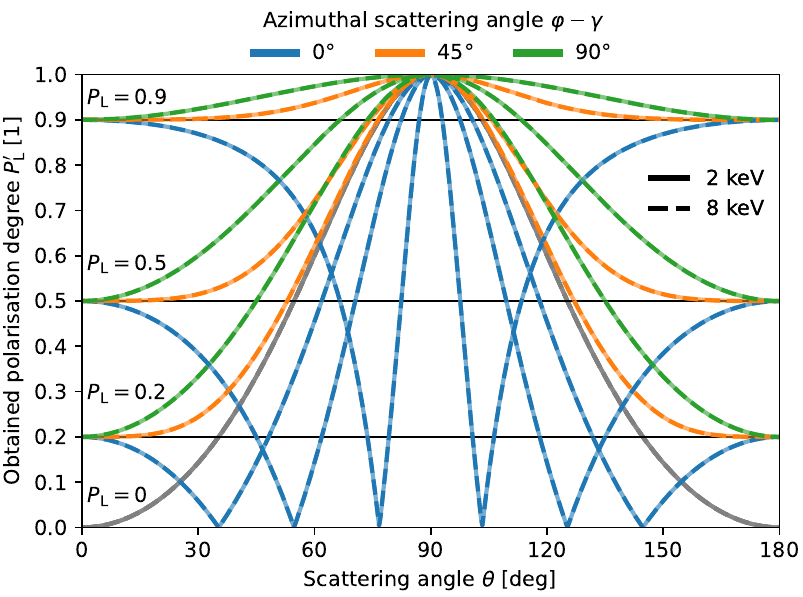}
    \caption{Linear polarisation degree $P'_{\text{L}}$ induced by Compton scattering, for photons with an initial polarisation degree $P_{\text{L}}$ of $0$, $0.2$, $0.5$ and $0.9$. $P'_{\text{L}}$ varies strongly with the scattering direction (with $\varphi-\gamma$ being the azimuthal scattering angle relative to the initial polarisation direction). For $P_{\text{L}}=0$ (grey), $P'_{\text{L}}$ is independent of the azimuthal direction by definition. The energy dependence in the $2-8~\text{keV}$ range is negligible.}
    \label{fig:PL}
\end{figure}
Through Compton scattering, the radiation field obtains a linear polarisation degree $P'_{\text{L}}$, which depends on the initial polarisation state ($P_{\text{L}}$ and $\gamma$) and the scattering direction ($\theta$ and $\varphi$):
\begin{equation}
    P'_{\text{L}} = \frac{\sqrt{\left(  {\mathrm{S}}_{12}  + {\mathrm{S}}_{22} P_{\text{L}} \cos2(\gamma-\varphi)\right)^2 + \left({\mathrm{S}}_{33} P_{\text{L}} \sin2(\gamma-\varphi)\right)^2}}{{\mathrm{S}}_{11} + {\mathrm{S}}_{12}P_{\text{L}} \cos2(\gamma-\varphi)}, \label{eq:PLprime}
\end{equation}
combing Eq.~(\ref{eq:PL}), (\ref{eq:QUinterpretation}) and (\ref{eq:updateStokes}), with ${\mathrm{S}}_{ij}\equiv {\mathrm{S}}_{ij}(\theta, x)$ being the M\"uller matrix elements given by Eq.~(\ref{eq:matrixelements}). For various incident polarisation degrees $P_{\text{L}}$, the resulting polarisation degree $P'_{\text{L}}$ is shown in Fig.~\ref{fig:PL}, illustrating how any polarisation degree between $0$ and $1$ can be obtained, depending on the specific scattering direction. Regardless of the initial polarisation state, scattered photons are maximally polarised ($P'_{\text{L}}=1$) for $\theta=90^\circ$, while forward scattering and backscattering leaves the polarisation degree unchanged. The resulting polarisation degree is shown at $2~\text{keV}$ (solid line) and $8~\text{keV}$ (dashed line), demonstrating how the energy dependence of $P'_{\text{L}}$ is negligible in the IXPE range. 

Equivalently, the linear polarisation angle obtained through Compton scattering is
\begin{equation}
    \gamma' = \frac12\arctan_2\left(\frac{{\mathrm{S}}_{33}(\theta, x)\, P_{\text{L}} \sin2(\gamma-\varphi)}{{\mathrm{S}}_{12}(\theta, x) + {\mathrm{S}}_{22}(\theta, x)\, P_{\text{L}} \cos2(\gamma-\varphi)}\right),
    \label{eq:gammaprime}
\end{equation}
combing Eq.~(\ref{eq:gamma}), (\ref{eq:QUinterpretation}) and (\ref{eq:updateStokes}). However, we remind that $\gamma'$ as given by Eq.~(\ref{eq:gammaprime}) refers to a different reference direction for each pair of scattering angles ($\theta, \varphi$) (see Eq.~(\ref{eq:dprime})). Therefore, these polarisation angles cannot be directly compared, except for some special symmetric cases such as the scenario discussed in Sect.~\ref{sec:poldegree}. For unpolarised photons ($P_{\text{L}}=0$), we recover the standard result that scattering induces polarisation perpendicular to the scattering plane ($\gamma'=90^\circ$, with $\bfd'$ in the scattering plane). 

\section{SKIRT implementation}
\label{sec:implementation}
\subsection{The radiative transfer code SKIRT}
\label{sec:skirt}
For the remainder of this work, we focus on radiative transfer simulations with the radiative transfer code SKIRT \citep{baes03, baes11, camps15a, camps20}. SKIRT is a state-of-the-art Monte Carlo radiative transfer code, developed and maintained at Ghent University, which implements a Monte Carlo photon life cycle emulating absorption, scattering, and re-emission in complex 3D transfer media \citep[see][for a review of the Monte Carlo radiative transfer method]{noebauer19}. SKIRT simulations in full 3D are facilitated by the implementation of various acceleration techniques \citep{baes08,steinacker13,baes16,baes22}, advanced grids for discretising the transfer medium \citep{saftly13, saftly14, camps13, lauwers24}, and a hybrid parallelisation scheme which combines multi-threading with multi-processing \citep{verstocken17,camps20}. The SKIRT code is open-source, well-documented\footnote{\url{https://skirt.ugent.be}}, and publicly available online\footnote{\url{https://github.com/SKIRT/SKIRT9}}, with tutorials for users and developers.

SKIRT models absorption, scattering, and re-emission in dusty astrophysical systems \citep{camps15a}, which includes emission from stochastically heated dust grains \citep{camps15b}, polarisation due to scattering on spherical dust grains \citep{peest17}, and polarised emission from aligned spheroidal dust grains \citep{vandenbroucke21}. Beyond dust, SKIRT models scattering on free electrons \citep{peest17, vandermeulen23}, absorption and emission at the $21~\text{cm}$ line of neutral hydrogen \citep{gebek23}, non-LTE line radiative transfer in the (sub)mm and infrared \citep{matsumoto23}, and resonant line scattering of H Ly$\alpha$ photons \citep{camps21}. Recently, the SKIRT code was extended into the X-ray regime, to model Compton scattering on free electrons, photo-absorption and fluorescence by cold atomic gas, scattering on bound electrons, and extinction by dust \citep{vandermeulen23}. Furthermore, the kinematics of moving sources and moving transfer media are self-consistently incorporated into the SKIRT radiative transfer calculations, so that the effect of bulk velocities and velocity dispersions can be properly modelled \citep{camps20}.

The SKIRT code features a large suite of  model geometries, radiation sources, medium characterisations, instruments, and probes \citep{baes08, camps15a, baes15}, in addition to interfaces for post-processing hydrodynamical simulations \citep{camps16, trcka20}. SKIRT can calculate self-consistent fluxes, images, spectra and polarisation maps from mm to X-ray wavelengths, with recent applications in  galaxies \citep[e.g.][]{hsu23, jang23, barrientos23,  savchenko23, kapoor23}, active galactic nuclei \citep[e.g.][]{stalevski23, kakkad23, gonzalez23, isbell23}, and other fields of astrophysics \citep[e.g.][]{jaquez23}.

\subsection{X-ray polarisation implementation}
\label{sec:notes}
The polarisation framework for tracing Stokes vectors in SKIRT has been implemented by \citet{peest17}, while the X-ray physical processes in cold neutral media (without polarisation support) have been implemented by \citet{vandermeulen23}. As photo-absorption and fluorescence do not depend on the polarisation state of the incoming photon, the original implementation is retained for these processes, with the important distinction that the Stokes vector is reset after fluorescent re-emission (i.e.\ the $Q'$, $U'$, and $V'$ parameters are set to zero; see Sect.~\ref{sec:processes}).

In this work, bound-electron scattering in cold atomic gas is approximated as free-electron scattering on $Z$ free electrons per neutral gas atom, as discussed in Sect.~\ref{sec:processes}. For a gas column with column density $N_\text{H}$, the optical depth for scattering is thus
\begin{equation}
    \tau(E) = \left(\sum_{Z} \: a_Z\, Z\right) \cdot N_\text{H} \cdot \sigma(E) ,
    \label{eq:tau}
\end{equation}
with $a_Z$ being the number density of element $Z$ relative to H{\,\textsc{i}}, and $\sigma(E)$ the Compton scattering cross-section given by Eq.~(\ref{eq:totalcompton}).
As this optical depth does not depend on the polarisation state of the incoming photon, sampling for scattering positions is conceptually identical to the free-electron scattering implementation as presented in \citet{vandermeulen23}. Given the abundance table $\left\{ a_Z\right\}$, one can easily generate a random interaction point in the forward $\bfk$ direction, following for example \citet[][]{whitney11}. 

Once the scattering location is set, a pair of scattering angles $\left(\theta, \varphi\right)$ can be sampled from the phase function Eq.~(\ref{eq:Comptonphasefunction}) using the conditional distribution method. Due to its symmetry around $\Psi(\theta, x)$, the azimuthal modulation term does not contribute to the marginal probability distribution for $\theta$, and therefore, random scattering angles $\theta$ can be generated from the same univariate distribution $ p(\theta; x)$ as for unpolarised Compton scattering:
\begin{align}
    p(\theta; x) 
    &= \frac{3\, \sigma_\text{T} \, {\mathrm{S}}_{11}(\theta, x)\sin\theta}{8\, \sigma(x)}.
\end{align}
This distribution for $\theta$ is a complex function of the incoming photon energy, but can be sampled efficiently using a variation on Khan's technique described by \citet{hua97}. Once a scattering angle $\theta$ is generated, we obtain the conditional distribution for the azimuthal scattering angle $\varphi$ as
\begin{align}
\label{eq:conditionalphi}
p_\theta(\varphi; x, {\textbf{S}}) 
&= \frac{1}{2\pi}\left( 1 + \frac{{{\mathrm{S}}_{12}(\theta, x)}}{{\mathrm{S}}_{11}(\theta, x)}P_\text{L}\cos2(\varphi-\gamma)\right),
\end{align}
which explicitly depends on the incoming polarisation state. A random scattering angle $\varphi$ can be sampled from Eq.~(\ref{eq:conditionalphi}) through numerical inversion as
\begin{align}
\chi &= \int_0^\varphi p_\theta(\varphi'; x, {\textbf{S}})\, d\varphi'\notag\\
&= \frac{1}{2\pi}\left( \varphi + \frac{{{\mathrm{S}}_{12}(\theta, x)}}{{\mathrm{S}}_{11}(\theta, x)}P_\text{L}\sin\varphi\cos(\varphi-2\gamma)\right),
\end{align}
with $\chi$ being a uniform deviate between 0 and 1.

Once the scattering angles $\theta$ and $\varphi$ are generated, the photon energy of the interacting photon is updated to $E'=C(\theta, x) \times E$ and the Stokes parameters are updated as described by Eq.~(\ref{eq:updateStokes}) (normalised so that $I'=C(\theta, x) \times I$, as discussed in Sect.~\ref{sec:compton}). Hereafter, the scattered photon continues its way through the transfer medium, in a direction $\bfk'$ that can be calculated from the original direction $\bfk$ and the scattering angles $\left(\theta, \varphi\right)$.

Eventually, all photons packets are recorded by (a set of) SKIRT instruments (corresponding to preset observer directions), using the peel-off technique as described in \citet{baes11}. We remind that the Stokes vectors of recorded photons are first rotated to correspond to the observer's north direction (following the IAU conventions; see Sect.~\ref{sec:stokes}), before they are binned to produce Stokes spectra and polarisation maps.

\subsection{Implementation verification}
\label{sec:verification}
\subsubsection{Simulation setup}
\label{sec:verificationsetup}
\begin{figure}
    \centering
	\includegraphics[width=0.8\columnwidth, trim={5.5cm 6cm 12cm 13cm},clip]{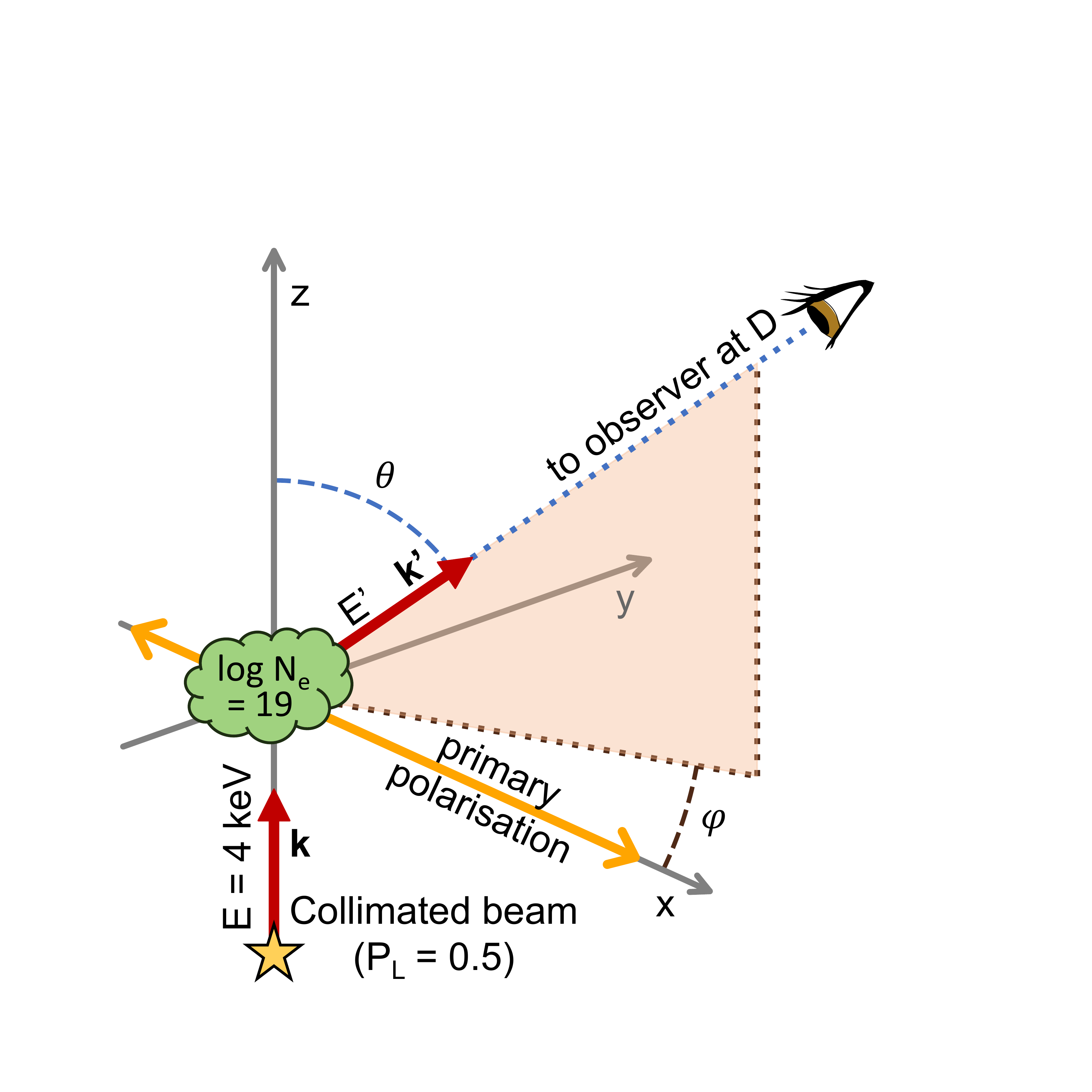}
    \caption{Three-dimensional setup for the analytical test cases in Sect.~\ref{sec:verification}: The primary source photons ($50\%$ linearly polarised along the $x$-axis) are emitted in the positive $z$-direction $\bfk$ and observed in a direction $\bfk'$ after Compton scattering on free electrons at the origin. The photons are emitted at $E = 4~\text{keV}$ and observed at $E' = C(\theta, x) \times E$ (see Eq.~(\ref{comptonfactor})). Having $N_\text{e} = 10^{19}~\text{cm}^{-2}$, the effect of multiple scattering is negligible.}
    \label{fig:verification}
\end{figure}
To verify the X-ray polarisation implementation in SKIRT, we set up dedicated SKIRT simulations, specifically designed to recover the details of individual photon-electron interactions. The simulation setup is shown in Fig.~\ref{fig:verification} and consists of a collimated beam illuminating a cloud of free electrons at the origin. The $4~\text{keV}$ source photons are emitted in the positive $z$-direction~$\bfk$ and are linearly polarised along the $x$-axis, with $P_{\text{L}}=0.5$. Without loss of generality, we can choose $\bfd=\boldsymbol{e}_{x}$ as the reference direction (with $\boldsymbol{e}_{x}$ being the Cartesian unit vector in the positive $x$-direction), so that $\gamma = 0$ by definition. The central electron cloud is small compared to its distance to the source and has a column density of $10^{19}~\text{cm}^{-2}$, making the flux contribution of photons that scatter more than once negligible (i.e.\ $<0.001\%$ of the one-time-scattered flux). The distance to the observer is $D$, and the spherical coordinates of the observer direction $\bfk'$ are just the scattering angle $\theta$ and the azimuthal scattering angle $\varphi$.

For any observer direction $\bfk'$, the scattered photon flux can be obtained with SKIRT, and because of the peel-off detection technique, this is possible for any exact ($\theta, \varphi$)-direction, with no spurious blurring due to averaging over some finite solid angle around $\bfk'$ \citep{yusefzadeh84}. Furthermore, the `smart' photon detectors in SKIRT can recover the individual flux contributions of direct (i.e.\ non-interacting) and reprocessed photons \citep{baes08}, in addition to recording the total Stokes $I$, $Q$, $U$, and $V$ spectra. For all (non-forward) directions, the total observed flux is equal to the scattered flux component, which is virtually equal to the one-time-scattered flux  (as further interactions are negligible at $N_\text{e} = 10^{19}~\text{cm}^{-2}$).

As a first sanity check, we calculated the ratio of the observed direct spectrum over the input spectrum for the forward direction ($\theta=0$), recovering the expected result of $1-\exp{\left[{-N_\text{e}\cdot \sigma(E)}\right]}$, with $\sigma(E)$ being the scattering cross-section given by Eq.~(\ref{eq:totalcompton}). In the next subsections, we use the SKIRT simulation output to recover the phase function (Sect.~\ref{sec:phasefunction}), the polarisation degree (Sect.~\ref{sec:poldegree}), and the polarisation angle (Sect.~\ref{sec:polangle}), which are then verified against the analytical formula obtained in Sect.~\ref{sec:compton}.

\subsubsection{Phase function}
\label{sec:phasefunction}
The scattering phase function $\Phi$ for polarised photons ($P_{\text{L}}>0$) depends explicitly on the azimuthal scattering angle $\varphi$, which is not the case for unpolarised Compton scattering (see Eq.~(\ref{eq:Comptonphasefunction})). Using the SKIRT output for the simulation setup described in Sect.~\ref{sec:verificationsetup}, we can infer the scattering phase function $\Phi_\mathrm{SKIRT}$ as implemented in SKIRT by relating the (one-time) scattered photon flux density $I(\theta, \varphi)$ in a direction $\bfk'$ to the fraction of the specific beam luminosity $L$ that interacted inside the electron cloud:
\begin{equation}
    \Phi_\mathrm{SKIRT}(\theta, \varphi; E, \bfS)= \frac{I(\theta, \varphi; E')\cdot C(\theta, E)^2 \cdot D^2}{\left(1-\exp{\left[{-N_\text{e}\cdot \sigma(E)}\right]}\right)\cdot L(E)},
    \label{eq:PhiSKIRT}
\end{equation}
with $I(\theta, \varphi)$ being the photon flux measured at $E' = C(\theta, E) \times E$, as Compton scattering is inelastic, and $L$ and $D$ being known input parameters. We note that the factor $C(\theta, E)^2$ appears in the nominator of Eq.~(\ref{eq:PhiSKIRT}) as the interval $\text{d}E$ around $E$ corresponds to the interval $\text{d}E'=C(\theta, E)^2\times \text{d}E$ around $E'$ after scattering.

\begin{figure}
    \centering
	\includegraphics[width=1\columnwidth, trim={0cm 0cm 0cm 0cm},clip]{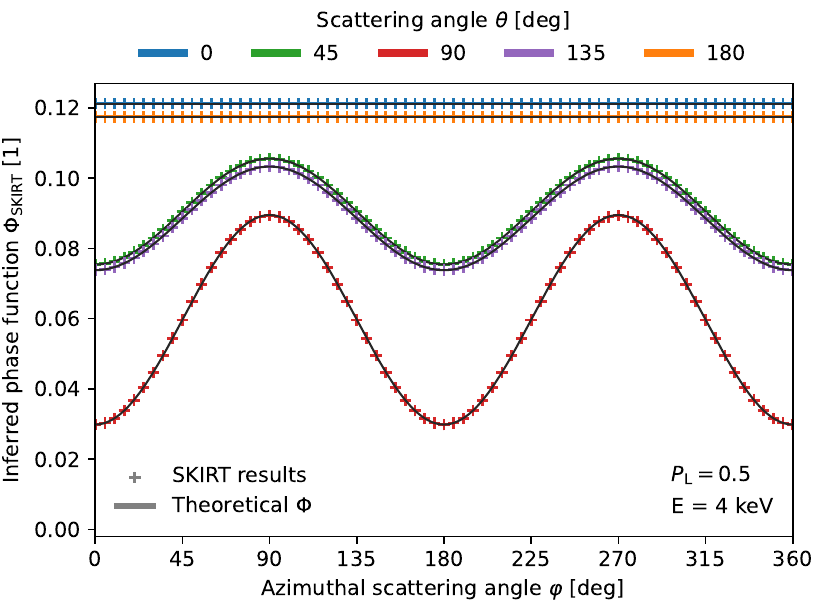}
    \caption{Compton phase function inferred from the simulation results for the three-dimensional setup described in Sect.~\ref{sec:verificationsetup} (see Fig.~\ref{fig:verification}). We find a perfect agreement between the SKIRT results and the theoretical phase function Eq.~(\ref{eq:Comptonphasefunction}), for a wide range of scattering angles ($\theta, \varphi$). }
    \label{fig:PhaseFunctionVerification}
\end{figure}
Fig.~\ref{fig:PhaseFunctionVerification} shows the scattering phase function $\Phi_\mathrm{SKIRT}$, which was inferred from the SKIRT results for the simulation setup described in Sect.~\ref{sec:verificationsetup}. Comparing this $\Phi_\mathrm{SKIRT}$ to the theoretical phase function given by Eq.~(\ref{eq:Comptonphasefunction}), we find an excellent agreement for all scattering angles ($\theta, \varphi$), recovering the prominent azimuthal modulation that governs the scattering physics for polarised photons. Using Eq.~(\ref{eq:PhiSKIRT}), we verified the SKIRT results for a range of polarisation states $\bfS$ and photon energies $E$ (especially beyond the IXPE range, where the energy dependence becomes more pronounced), assuring a correct implementation of polarisation-dependent Compton scattering in SKIRT.

\subsubsection{Polarisation degree}
\label{sec:poldegree}
\begin{figure}
    \centering
	\includegraphics[width=1\columnwidth, trim={0cm 0cm 0cm 0cm},clip]{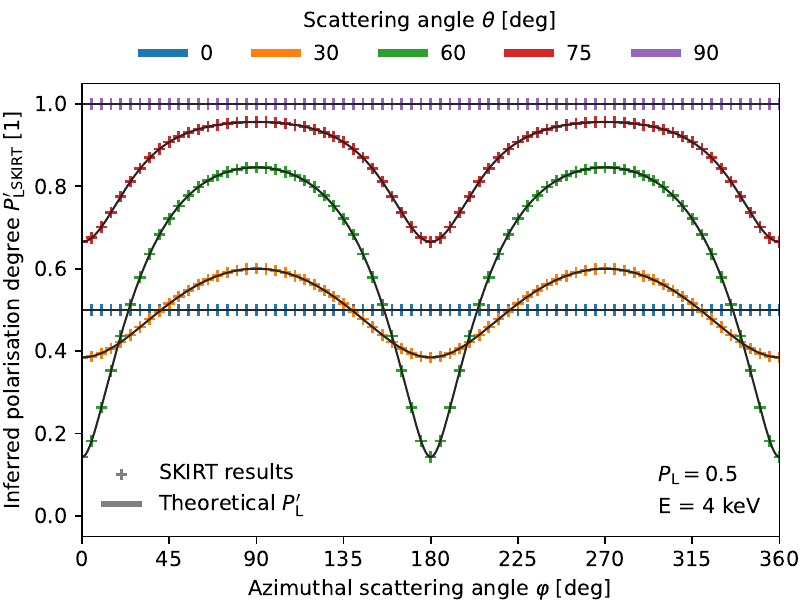}
    \caption{Linear polarisation degree ${P_{\mathrm{L}}'}$ inferred from the simulation results, for the 3D setup described in Sect.~\ref{sec:verificationsetup} (see Fig.~\ref{fig:verification}). We find an excellent agreement between the SKIRT results and the theoretical polarisation degree Eq.~(\ref{eq:PLprime}), for all scattering angles ($\theta, \varphi$).}
    \label{fig:PolDegVerification}
\end{figure}
The Stokes $I(\theta, \varphi)$, $Q(\theta, \varphi)$, and $U(\theta, \varphi)$ spectra calculated with SKIRT can be used to obtain the polarisation degree for the setup described in Sect.~\ref{sec:verificationsetup} by applying Eq.~(\ref{eq:PL}). Fig.~\ref{fig:PolDegVerification} shows the inferred polarisation degree as a function of the scattering angle, together with the theoretical formula for ${P_{\mathrm{L}}'}$, given by Eq.~(\ref{eq:PLprime}). As the polarisation degree induced through Compton scattering is virtually symmetric around $\theta=90^\circ$ (see Fig.~\ref{fig:PL}), we focus on scattering angles $\theta\leq 90^\circ$ only. We find an excellent agreement between the SKIRT results and the theoretical formula Eq.~(\ref{eq:PLprime}), recovering the complex behaviour of ${P_{\mathrm{L}}'}$ with $\theta$ and $\varphi$. We find that forward scattering ($\theta= 0$) does not change the polarisation degree, while ${P_{\mathrm{L}}'}=1$ when $\theta=90^\circ$, as predicted by Fig.~\ref{fig:PL}.

\subsubsection{Polarisation angle}
\label{sec:polangle}
\begin{figure}
    \centering
	\includegraphics[width=1\columnwidth, trim={0cm 0cm 0cm 0cm},clip]{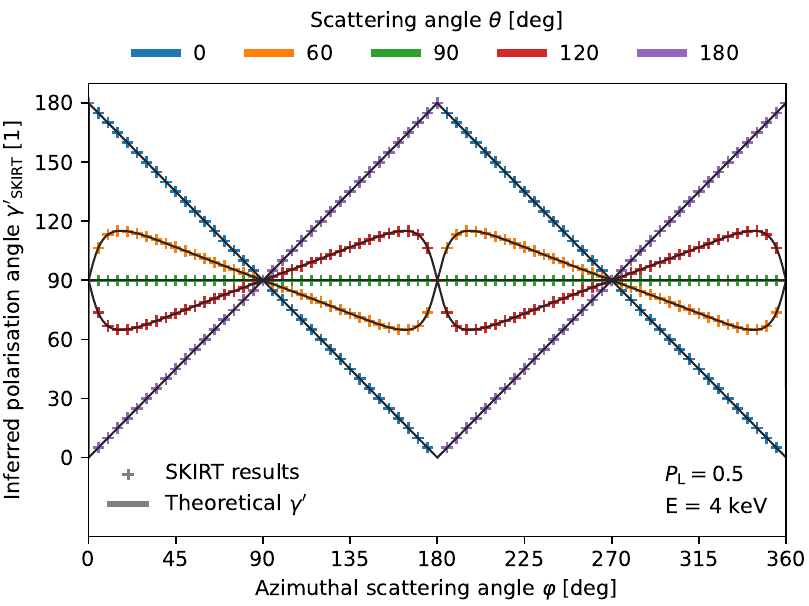}
    \caption{Linear polarisation angle $\gamma'$ inferred from the simulation results, showing an excellent agreement between the SKIRT output and the theoretical formula Eq.~(\ref{eq:gammaprime}), for all scattering angles ($\theta, \varphi$).}
    \label{fig:PolAngVerification}
\end{figure}
Similarly, the linear polarisation angle ${\gamma'}$ can be calculated from the Stokes spectra obtained with SKIRT by applying Eq.~(\ref{eq:gamma}). The inferred polarisation angle $\gamma'_\mathrm{SKIRT}$ is then defined relative to the north direction of the observer frame, as discussed in Sect.~\ref{sec:notes}. This $\gamma'_\mathrm{SKIRT}$ can be directly compared to the theoretical formula as given by Eq.~(\ref{eq:gammaprime}), as for this particular setup (Sect.~\ref{sec:verificationsetup}), Eq.~(\ref{eq:gammaprime}) refers to a reference direction $\bfd'$ that is also the observer north direction\footnote{Using Eq.~(\ref{eq:dprime}) with $\bfk=\boldsymbol{e}_{z}$ and $\bfd=\boldsymbol{e}_{x}$ (see Sect.~\ref{sec:verificationsetup}), $\bfd'$ is just equal to the spherical unit vector $\boldsymbol{e}_{\theta}$. The observer north direction is $-\boldsymbol{e}_{\theta}$, but Stokes vectors are invariant under rotations over $180^\circ$ (see Eq.~(\ref{eq:rotation})).}.

The linear polarisation angle inferred from the SKIRT output is shown in Fig.~\ref{fig:PolAngVerification}, for various scattering angles. For forward scattering and backscattering ($\theta= 0$ and $180^\circ$, respectively), the observed $\gamma'$ is just the projected\footnote{Projected on the sky as seen from the observer location.} incident polarisation direction, as the polarisation direction is retained by scattering at $\theta = 0^\circ$ or $\theta = 180^\circ$ (see Eq.~(\ref{eq:matrixelements}) and~(\ref{eq:updateStokes})). For intermediate scattering angles, the observed polarisation angle lays between ${\gamma'}=90^\circ$ (i.e.\ perpendicular to the scattering plane) and the specific angle corresponding to the projected incoming polarisation direction. For $\varphi = 90^\circ$ or $270^\circ$, the projected incident polarisation direction is zero, and therefore, the observed polarisation direction is just perpendicular to the scattering plane (${\gamma'}=90^\circ$),  identical to the case of unpolarised incident photons (see Sect.~\ref{sec:compton}). 

We find an excellent agreement between the SKIRT results and the theoretical formula Eq.~(\ref{eq:gammaprime}) for ${\gamma'}$, recovering the complex behaviour of the polarisation angle with the scattering direction. Together with the results of Sect.~\ref{sec:phasefunction} and Sect.~\ref{sec:poldegree}, this assures us that the Compton scattering formula are implemented correctly in SKIRT.

\section{Torus models}
\label{sec:torus}
\subsection{Model setup}
\label{sec:setup}
As a first application of the new X-ray polarisation capabilities of the SKIRT code, we explored the spectro-polarimetric properties of a 2D toroidal reprocessor of cold gas, modelling the parsec-scale circumnuclear medium of AGN. This medium is expected to reprocess the primary X-ray emission of the central X-ray corona, producing a distinct polarisation signal which can be used to constrain the geometry of the reprocessor in heavily obscured AGN \citep[see e.g.][]{ursini23}. 

\begin{figure}
    \centering
	\includegraphics[width=1\columnwidth, trim={1cm 4.4cm 0 3.5cm},clip]{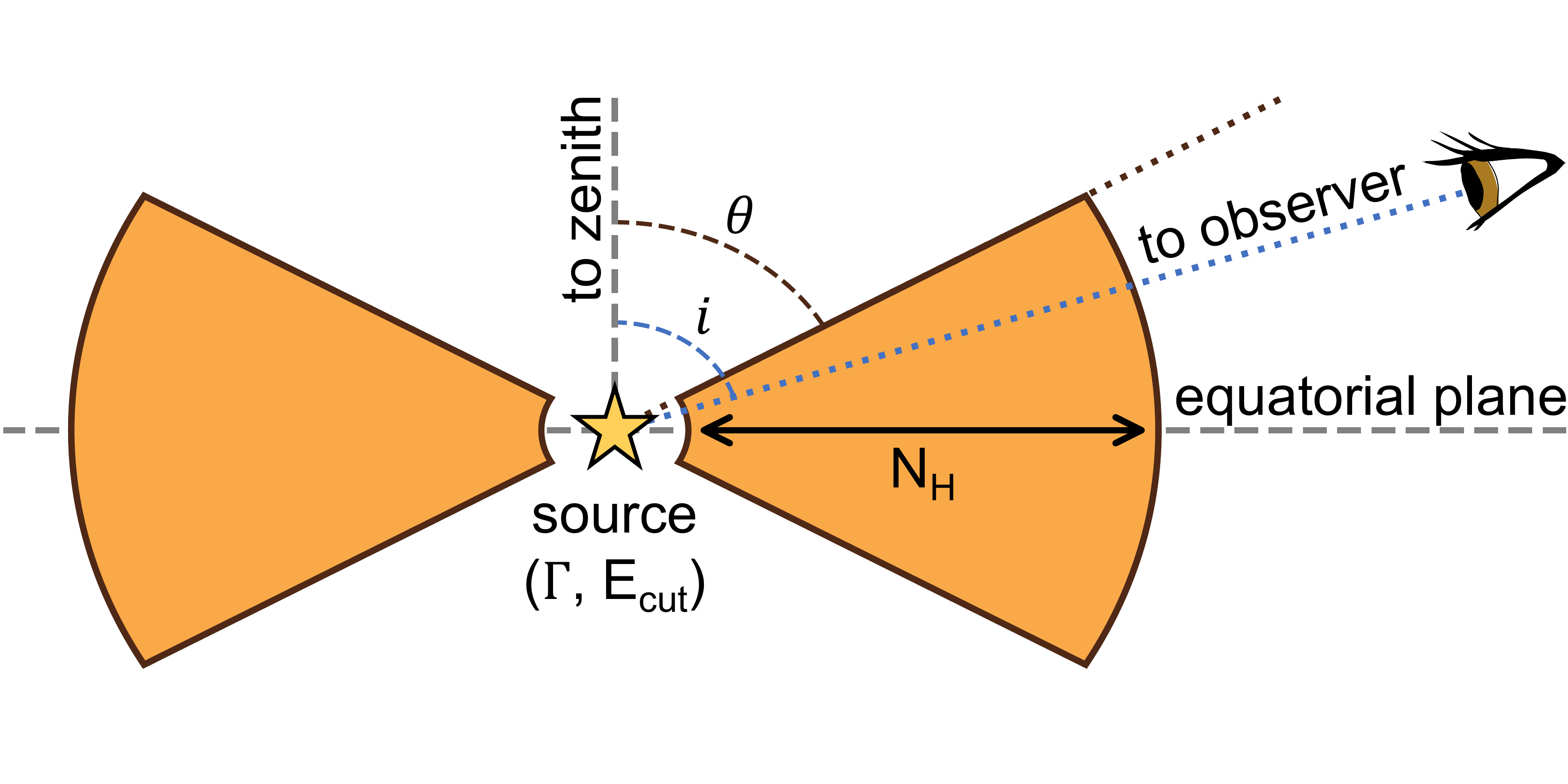}
    \caption{Wedge torus geometry modelling the circumnuclear medium in Sect.~\ref{sec:torus}. We adopted a uniform distribution of cold gas, with a covering factor $\text{CF}=\cos\theta$ and an equatorial hydrogen column density $N_\text{H}$, which is observed at an inclination $i$. The (unpolarised) central X-ray source has a power law spectrum with a photon index $\Gamma$ and a cut-off at $E_\textrm{cut}$.}
    \label{fig:geometry}
\end{figure}
We adopted a wedge torus geometry of uniform-density gas, centred on an isotropic point source representing the X-ray corona (see Fig.~\ref{fig:geometry}). This torus geometry is identical to the torus geometries of the \textsc{BNTorus} \citep{brightman11} and \textsc{borus} \citep{balokovic18, balokovic19} spectral models and is also known as a `flared disc'. The X-ray radiative transfer problem is scale invariant, and therefore the model geometry is fully defined by the torus covering factor $\text{CF} = \cos{\theta}$ and the equatorial hydrogen column density $N_\text{H}$, which are indicated on Fig.~\ref{fig:geometry}. For the wedge torus geometry, all obscured sightlines ($\cos{i} < \text{CF}$) have the same line-of-sight $N_\text{H}$, while unobscured sightlines have $N_\text{H} = 0$.

We considered photo-absorption, fluorescence, and scattering by cold neutral gas as described in Sect.~\ref{sec:processes}, and  adopted \citet{anders89} solar gas abundances. We assumed a standard power law spectrum for the central X-ray source, characterised by a photon index $\Gamma$ and an exponential cut-off energy $E_\textrm{cut}$. The primary X-ray photons were assumed to be unpolarised (${P_{\mathrm{L}}}=0$), so that the emerging polarisation signal could be entirely attributed to reprocessing in the AGN torus. Indeed, IXPE observations of unobscured AGN indicate that the coronal polarisation levels are low \citep{marinucci22, gianolli23, ingram23}, so that the initial polarisation state of the coronal X-ray photons is easily washed out by multiple scattering in the AGN torus \citep{marin18a}.

We set up radiative transfer simulations in the wedge torus geometry, using the SKIRT code with X-ray polarisation enabled\footnote{SKIRT version 9 git commit \texttt{21bd407}.}. First, we specified the spatial grid on which the torus medium was discretised. In this case, the axial symmetry of the torus could be exploited to build a 2D grid, which would speed up the radiative transfer simulations significantly. Furthermore, the density distribution in the azimuthal plane could be gridded using a 2D polar grid with angular bins coinciding with the torus edges, so that the gridded distribution would equal the exact model distribution, with no discretisation effects.

Stokes spectra were calculated over the $0.3$ to $200~\text{keV}$ range, adopting $1000$ logarithmic wavelength bins, which corresponds to a spectral resolution of $R=154$. In SKIRT, X-ray radiative transfer interactions were modelled up to $500~\text{keV}$, as the X-ray reflection spectrum and the corresponding polarisation signal are produced by Compton down-scattering at energies beyond $200~\text{keV}$ \citep[see e.g.][for a discussion]{vandermeulen23}.

\subsection{Radiative transfer results}
\label{sec:results}
\begin{figure*}
    \centering
	\includegraphics[width=0.95\hsize]{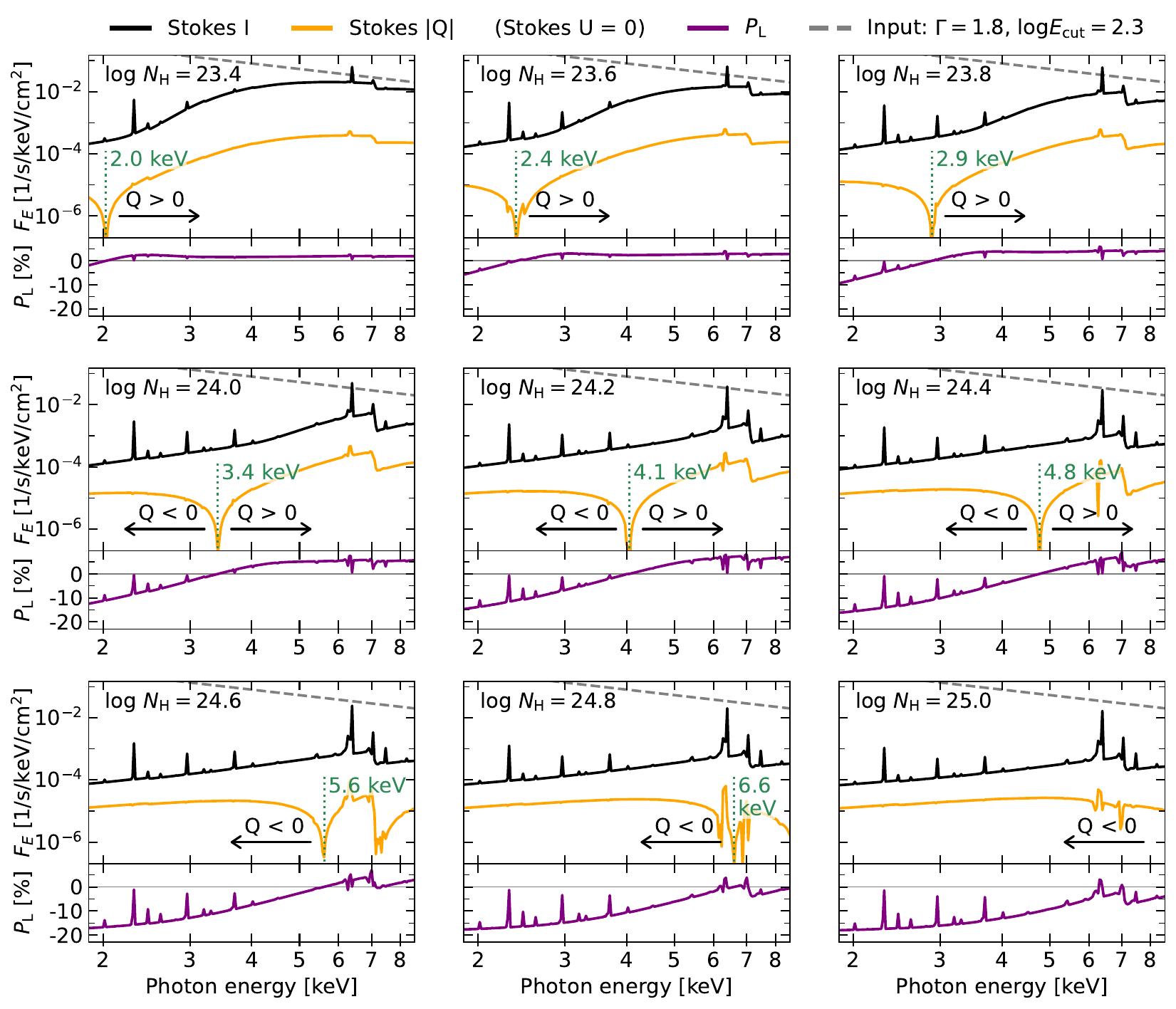}
    \caption{SKIRT radiative transfer results for the wedge torus model shown in Fig.~\ref{fig:geometry}, for a covering factor $\text{CF}=0.45$, $\cos{i}=0.4$, and source parameters representative for Circinus AGN ($\Gamma = 1.8$,  $E_\textrm{cut}=200~\text{keV}$ and $L_{2-10}=2.8 \times 10^{42}~\text{erg}~\text{s}^{-1}$). The different panels represent different values for the equatorial hydrogen column density $N_\text{H}$. Because of the torus symmetry, $U$ is always zero. Positive values for the linear polarisation degree $P_{\text{L}}$ (purple line) denote polarisation in the north direction of the observer frame ($\gamma=0^\circ$), which is the direction of the torus symmetry axis projected on the sky. Negative $P_{\text{L}}$ values denote polarisation perpendicular to this projected symmetry axis, which is the horizontal direction in the observer frame ($\gamma=90^\circ$). The photon energy $E_\text{flip}$ where the polarisation angle $\gamma$ flips from $90^\circ$ to $0^\circ$ is indicated in green on all panels.}
    \label{fig:spectra}
\end{figure*}
The SKIRT radiative transfer results are shown in Fig.~\ref{fig:spectra} for one particular realisation of the wedge torus model ($\text{CF}=0.45$ and $\cos{i}=0.4$), with different panels representing different values for the column density, ranging from $\log{N_\text{H}}=23.4$ (Compton-thin, top left) to $\log{N_\text{H}}=25.0$ (Compton-thick, bottom right). The adopted source parameters are representative for the nearby AGN in the Circinus galaxy ($\Gamma = 1.8$,  $E_\textrm{cut}=200~\text{keV}$, and $L_{2-10}=2.8 \times 10^{42}~\text{erg}~\text{s}^{-1}$, observed at a distance of $4.2~\text{Mpc}$). We focus on the SKIRT results in the $2-8~\text{keV}$ IXPE range.

The Stokes $I$, $Q$, and $U$ spectra are shown at the top of each panel. The observed Stokes $U$ parameter is always zero, which was expected as the north direction of the observer frame was chosen to just coincide with the projected symmetry axis of the torus (while further considering that a 2D system cannot exhibit features that are asymmetric relative to its projected symmetry axis). In a realistic observational context, the observed system will be rotated about the line of sight of the observer, mixing the $Q$ and $U$ parameters as described by Eq.~(\ref{eq:rotation}). This rotation-induced mixing of the Stokes parameters can be observed, which could be used to infer the orientation of the principal axis of the AGN system, forming a powerful probe on the (spatially unresolved) circumnuclear medium \citep[see e.g.][]{ursini23, marin24}.

From these Stokes spectra, the linear polarisation degree $P_{\text{L}}$ and the linear polarisation angle $\gamma$ were calculated as a function of the photon energy, using Eq.~(\ref{eq:PL}) and~(\ref{eq:gamma}). As Stokes $U$ is always zero, $\gamma$ can only be $0^\circ$ or $90^\circ$. Therefore, one could visualise the observed polarisation direction with positive $P_{\text{L}}$ values denoting polarisation in the north direction ($\gamma=0^\circ$) and negative $P_{\text{L}}$ values denoting polarisation in the horizontal direction ($\gamma=90^\circ$) (see $P_{\text{L}}$, in purple, at the bottom of each panel in Fig.~\ref{fig:spectra}). As the north direction of the observer frame was chosen to just coincide with the projected torus axis, $\gamma=0^\circ$ and $\gamma=90^\circ$ correspond to polarisation parallel and perpendicular to the torus symmetry axis, respectively.

The radiative transfer results shown in Fig.~\ref{fig:spectra} demonstrate the complex behaviour of the polarisation observables of a 2D torus model in the $2-8~\text{keV}$ range. Most prominently, we note that the Stokes $Q$ spectra have a fixed negative sign up to a certain photon energy $E_\text{flip}$ (indicated in green on Fig.~\ref{fig:spectra}), beyond which $Q$ becomes positive. With Stokes $U$ being zero, this means that for $E<E_\text{flip}$, the observed emission is polarised perpendicular to the torus symmetry axis (i.e.\ $\gamma=90^\circ$), while at higher energies, the polarisation is parallel to this symmetry axis (i.e.\ $\gamma=0^\circ$).

The exact photon energy $E_\text{flip}$ where the polarisation angle $\gamma$ flips from $90^\circ$ to $0^\circ$ is a function of the opacity of the torus medium, increasing from $2.0~\text{keV}$ at $\log{N_\text{H}}=23.4$ to $6.6~\text{keV}$ at $\log{N_\text{H}}=24.8$\footnote{For $\log{N_\text{H}}=25.0$, $E_\text{flip} = 10.4~\text{keV}$, outside of the IXPE range.}. Furthermore, the effect of the torus opacity is also observed in the $\log{N_\text{H}}=24.8$ panel of Fig.~\ref{fig:spectra}, where the polarisation angle $\gamma$ flips back to $90^\circ$ at $7.1~\text{keV}$  (i.e.\ at the Fe K absorption edge), where the torus opacity rises discontinuously. A second flip from $\gamma=90^\circ$ to $\gamma=0^\circ$ is then observed at $8.8~\text{keV}$. This behaviour of the polarisation angle with photon energy is discussed in Sect.~\ref{sec:polifoE}.

Away from $E_\text{flip}$, Stokes $Q$ is relatively featureless and roughly follows the spectral shape of the reflected continuum. In particular, the  polarised flux (i.e.\ Stokes $Q$) does not contain any fluorescent lines, as fluorescent line photons are emitted at $P_{\text{L}}=0$ (see Sect.~\ref{sec:processes}). However, as the total flux (i.e.\ Stokes $I$) contains strong fluorescent lines, the linear polarisation degree is heavily diluted at the fluorescent line energies, which produces line features where $P_{\text{L}}$ approaches zero (see Fig.~\ref{fig:spectra}). Similarly, at low $N_\text{H}$, the polarisation signal is heavily diluted by the direct flux of the (unpolarised) primary X-ray source.

\subsection{Model grid}
\label{sec:xspec}
\begin{table}
\caption{Free parameters of the torus model, with the adopted range of parameter values.}
\label{table:1}
\centering                                     
\begin{tabular}{c c c c}          
\hline\hline                       
Parameter & Minimum & Maximum &  Step\\    
\hline                                  
    $\log{N_\text{H}}$ & 22.0 & 26.0 & 0.2\\     
    $\Gamma$ & 1.0 & 3.0 & 0.2\\     
    $\log{E_\textrm{cut}}$ & 1.5 & 2.9 & 0.2\\     
    $\text{CF}$ & 0.05 & 0.95 & 0.1\\     
    $\cos{i}$ & 0.0 & 1.0 & 0.1\\     
\hline                                             
\end{tabular}
\tablefoot{The five parameters are the equatorial hydrogen column density ($\log{N_\text{H}}$), the photon index ($\Gamma$), the exponential cut-off energy ($\log{E_\textrm{cut}}$), the covering factor ($\text{CF} = \cos{\theta}$), and the inclination ($\cos{i}$). $N_\text{H}$ and $E_\textrm{cut}$ are expressed in units of $\textrm{cm}^{-2}$ and \textrm{keV}, respectively.}
\end{table}
We calculated a grid of AGN torus models for observational data fitting based on the wedge torus geometry described in Sect.~\ref{sec:setup} by varying $N_\text{H}$ between $10^{22}$ and $10^{26}~\text{cm}^{-2}$, the photon index $\Gamma$ between $1$ and $3$, and the exponential cut-off energy $\log E_\textrm{cut}$ between $1.5$ and $2.9$ (corresponding to $30~\text{keV}$ and $800~\text{keV}$, respectively). We considered torus covering factors between $0.05$ and $0.95$, observed at inclinations between $0^\circ$ and $90^\circ$, with all parameters being sampled as described in Table~\ref{table:1}. This leads to $203280$ unique torus model realisations, forming a parameter space that is covering most of the obscured AGN observed in the local Universe \citep{bianchi09, ricci15, ricci17, ricci18}. For each parameter combination, the Stokes $I$, $Q$, and $U$ spectra were calculated from $0.3$ to $200~\text{keV}$, with a spectral resolution of $R=154$ (see Sect.~\ref{sec:setup}).

For all model realisations, the number of simulated photon packets was kept at $10^8$, resulting in output spectra with limited Monte Carlo noise. The simulation run time of a single model realisation depends mostly on the covering factor and the column density of the torus, as more material requires more interactions to be modelled. In addition, the spectral hardness of the source has a small effect, as hard X-ray photons experience more scattering events before being absorbed. For a standard source spectrum with $\Gamma = 1.8$ and $E_\textrm{cut}=200~\text{keV}$, individual simulations take between $0.9$ and $33.1~\text{minutes}$ on a modest $2.2~\text{GHz}$ 16-core node, demonstrating the computational efficiency of the SKIRT code. By using the high-performance computing infrastructure at the Flemish Supercomputer Centre, it was possible to run all $203280$ torus model realisations in just $55~\text{hours}$.

The radiative transfer results were converted to an XSPEC \citep{arnaud96} table model named \textsc{xskirtor\textunderscore smooth}, which is publicly released with this work\footnote{\url{https://github.com/BertVdM/xskirtor}}. This model can be used for observational data fitting within XSPEC as
\begin{equation}
\label{eq:xspec}
    \textsc{model polrot $\times$ atable\{xskirtor\textunderscore smooth.mod\}}.
\end{equation}
In addition to the five free model parameters listed in Table~\ref{table:1}, the table model requires a redshift and a luminosity normalisation\footnote{Following the same convention as the \textsc{cutoffpl} model in XSPEC, the \textsc{norm} parameter is defined as the constant factor $K$ in the unobscured flux density of the primary source: $F_E(E) = K\; E^{-\Gamma} \exp\left(-E/E_\textrm{cut}\right)$, in units of counts/s/keV/cm$^2$. The \textsc{norm} parameter is approximately equal to the unobscured flux density at $1~\textrm{keV}$.}, providing the physical scaling of the model spectra. \textsc{polrot} then sets the roll angle $\theta$ of the torus system around the line of sight of the observer, as the table model corresponds to an observer that has its north direction coinciding with the projected torus symmetry axis. This brings the number of free model parameters to eight. The \textsc{xskirtor\textunderscore smooth} model can be further combined with other XSPEC models, such as a \textsc{tbabs} component to model galactic foreground extinction \citep{wilms00}.

The \textsc{xskirtor\textunderscore smooth} model predicts both X-ray spectra and X-ray spectro-polarimetry, enabling the simultaneous fitting of spectral and polarimetric observations in the X-ray band. As both the spectral coverage and the spectral resolution of the model templates are largely exceeding the capabilities of modern X-ray polarisation observatories, they can be applied for the interpretation of observational IXPE, XPoSat, and eXTP data, proposal writing, and the definition of future missions.

The corresponding spectral model (modelling Stokes $I$) is well suited for fitting CCD-based X-ray spectra as obtained with modern X-ray observatories such as \emph{XMM-Newton}, \emph{Chandra}, \emph{NuSTAR}, \emph{Swift}, INTEGRAL, \emph{AstroSat}, and more. Furthermore, we provided an additional XSPEC table model for the $1.5$ to $15~\text{keV}$ subrange, with an adaptive spectral resolution achieving $\Delta E = 0.5~\text{eV}$ around the strongest fluorescent lines, for fitting the high-resolution X-ray spectra that are being obtained with the XRISM/Resolve \citep[][]{tashiro20} and will be obtained with the \emph{Athena}/X-IFU \citep[][]{barret18}.

\section{Discussion}
\label{sec:discussion}
\subsection{Polarisation angle as a function of the photon energy}
\label{sec:polifoE}
\begin{figure*}
    \centering
	\includegraphics[width=0.95\hsize]{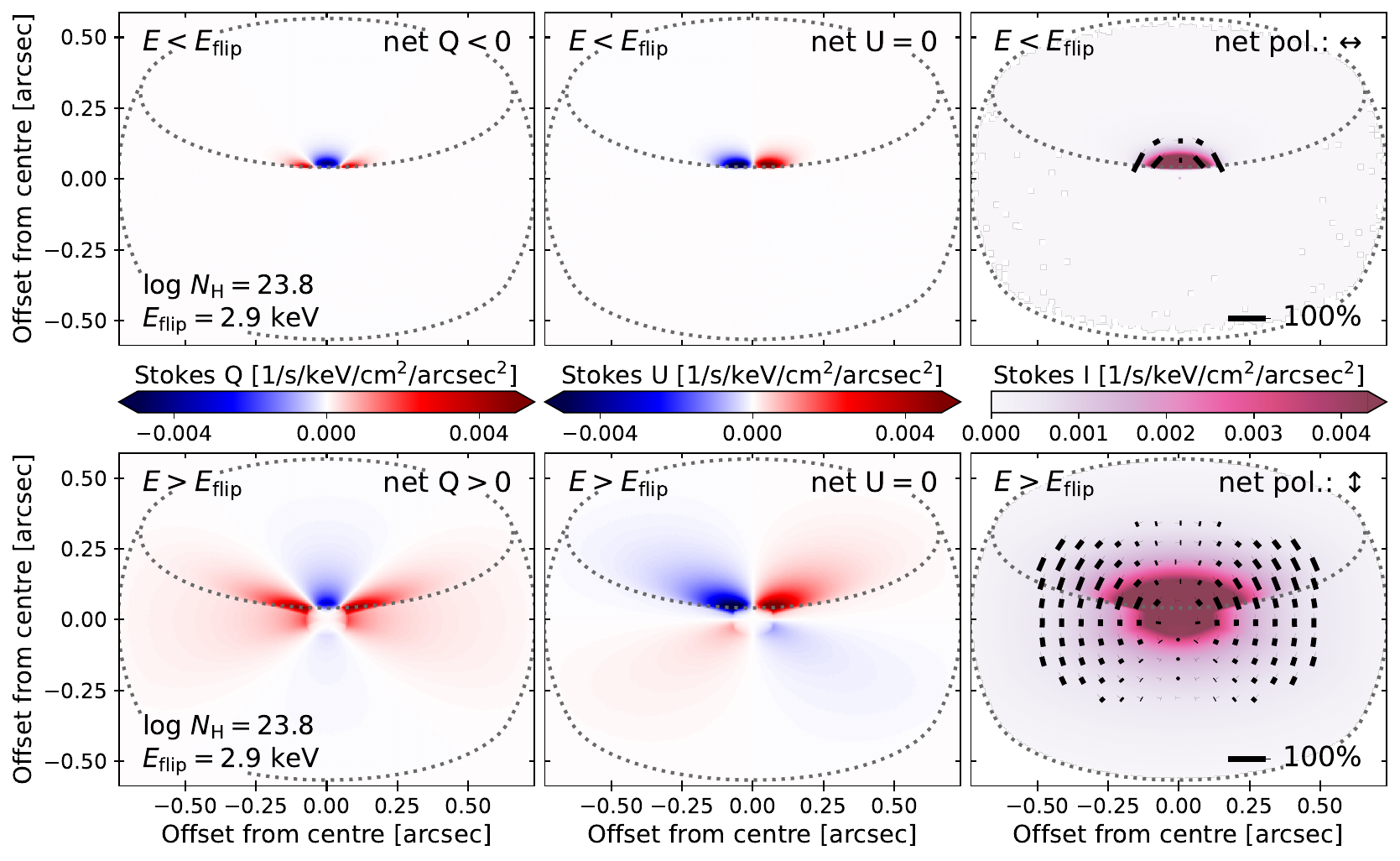}
    \caption{Spatially resolved Stokes surface brightness maps for the $\log{N_\text{H}}=23.8$ torus model realisation shown in the top right panel of Fig.~\ref{fig:spectra}. The top (bottom) row corresponds to the energy band below (above) $E_\text{flip}$, where the net polarisation signal is perpendicular (parallel) to the torus symmetry axis. The total surface brightness (Stokes $I$, shown on the right) is overlaid with a linear polarisation map, visualising the polarisation degree (length of the map segments) and polarisation angle (orientation of the map segments). The dotted grey lines indicate the torus outlines.}
    \label{fig:LinPolMap}
\end{figure*}
In Sect.~\ref{sec:results}, we described how the reprocessed torus emission is polarised perpendicular to the torus symmetry axis ($\gamma=90^\circ$) at low energies, while it is parallel to the torus symmetry axis ($\gamma=0^\circ$) at higher energies. This behaviour of the polarisation angle $\gamma$ with photon energy is observed for a wide range of torus parameter combinations (see Sect.~\ref{sec:setup}), with a transition energy $E_\text{flip}$ that scales with the torus opacity. To understand how this behaviour is related to the torus geometry, we can inspect the spatially resolved Stokes surface brightness maps, which can be calculated with SKIRT for each parameter combination.

As a demonstration, we focus on the $\log{N_\text{H}}=23.8$ torus model realisation that was discussed in Sect.~\ref{sec:results}, for which the Stokes spectra are shown in the top right panel of Fig.~\ref{fig:spectra}. The corresponding Stokes $Q$~(left), Stokes $U$~(middle), and Stokes $I$~(right) surface brightness maps are shown in Fig~\ref{fig:LinPolMap}, with the top row representing the $2~\text{keV}$ to $E_\text{flip}$ energy band, and the bottom row representing the $E_\text{flip}$ to $8~\text{keV}$ band, with $E_\text{flip} = 2.9~\textrm{keV}$. In addition, the total surface brightness map (Stokes $I$, on the right) is overlaid with a linear polarisation map, visualising the polarisation degree and polarisation angle as derived from the Stokes $Q$ and Stokes $U$ surface brightness maps.

Looking at the top right panel of Fig.~\ref{fig:LinPolMap}, we find that for $E<E_\text{flip}$, the total observed flux is dominated by reprocessed photons that scattered off the illuminated (unobscured) backside of the torus. For $E>E_\text{flip}$ on the other hand (bottom right), source photons are able to penetrate the obscuring front side of the torus, so that the total observed flux is dominated by direct source emission. However, this unpolarised direct flux merely dilutes the observed polarisation signal, without influencing the polarisation angle $\gamma$ that is eventually observed. Therefore, we can focus on the reprocessed flux only, finding that for $E>E_\text{flip}$, scattered photons originate from a more extended region on the illuminated backside of the torus, while they can also escape through the obscuring front side of the torus, as evident from the bottom right panel of Fig.~\ref{fig:LinPolMap}.

Using the smart photon detectors in SKIRT (see Sect.~\ref{sec:verification}), we can confirm that the reprocessed torus emission is dominated by one-time-scattered photons, which significantly simplifies the interpretation of the linear polarisation maps shown in Fig.~\ref{fig:LinPolMap}. As the primary source photons are unpolarised (see Sect.~\ref{sec:setup}), Compton scattering induces polarisation that is exactly perpendicular to the scattering plane (see Eq.~(\ref{eq:gammaprime})). This means that the projected polarisation pattern is always circular (i.e.\ in each pixel, the observed polarisation direction is perpendicular to the direction towards the central pixel), as shown in Fig.~\ref{fig:LinPolMap}. Indeed, the Stokes $Q$ surface brightness is positive in the east and west quadrants (describing polarisation in the vertical direction) and negative in the north and south quadrants (describing polarisation in the horizontal direction). Similarly, Stokes $U$ is positive in the north-east and south-west quadrants and negative in the north-west and south-east quadrants (see Fig.~\ref{fig:LinPolMap}). The polarisation direction of individual reprocessed photons thus depends on the specific sky region of the final scattering interaction inside the torus (in this specific case, dominated by single scattering). 

In addition, we find that also the observed polarisation degree depends on the sky region, with $P_{\text{L}}$ being high in the east and west quadrants and low in the north and south quadrants (see Fig.~\ref{fig:LinPolMap}, where $P_{\text{L}}$ is visualised by the length of the polarisation map segments). This behaviour is a direct result of the projected inclined torus geometry: In the north and south quadrants, the distribution of last scattering angles (i.e.\ to reach the observer) forms a narrow peak centred on $130^\circ$ and $50^\circ$, respectively. For both quadrants, this correspond to a distribution of polarisation degrees having $P_{\text{L}}<0.3$ for $60\%$ of all photons and $P_{\text{L}}<0.5$ for $90\%$ of all photons, as predicted by Eq.~(\ref{eq:PLprime}). On the other hand, the distribution of scattering angles is much broader in the east and west quadrants and peaks at $90^\circ$, which corresponds to higher average polarisation degrees (see Fig.~\ref{fig:PL}). Indeed, in these side quadrants, $50\%$ of all photons are having $P_{\text{L}}>0.6$.

Summarising, the net (i.e. spatially integrated) polarisation direction that is eventually observed, is the result of the precise balance between the polarised flux originating from the different sky regions, with both the total flux and the polarisation degree of each sky region being closely related to the torus geometry. As the projected torus geometry is perfectly symmetric around the observer north direction, the Stokes $U$ surface brightness maps are perfectly antisymmetric around this axis (see Fig.~\ref{fig:LinPolMap}), so that the spatially integrated Stokes $U$ fluxes are always zero, as discussed in Sect.~\ref{sec:results}. The Stokes $Q$ surface brightness maps on the other hand do not exhibit such trivial symmetry, so that the spatially integrated polarisation angle $\gamma$ depends on the balance between the polarised flux in the east and west quadrants (where Stokes $Q>0$ and $\gamma=0^\circ$) compared to the north and south quadrants (where Stokes $Q<0$ and $\gamma=90^\circ$).

We can now explain the behaviour of the polarisation angle with photon energy for this particular setup: At low energies ($E<E_\text{flip}$), the reprocessed torus emission is purely dominated by photons that scattered off a small region on the illuminated backside of the torus. This backside region mostly covers the northern sky quadrant where Stokes $Q$ is negative, so that the spatially integrated polarisation signal is perpendicular to the torus symmetry axis ($\gamma=90^\circ$). At higher energies ($E>E_\text{flip}$), the reprocessed torus emission originates from a more extended region on the torus backside and also escapes through the obscuring front side of the torus, so that all four sky quadrants are covered. At $E_\text{flip}$, the polarised flux is still dominated by the unobscured torus backside, but now the positive $Q$ contributions on the torus backside (close to the torus front edge, as shown in red) start to dominate over the negative $Q$ contributions (shown in blue; see the bottom left panel of Fig.~\ref{fig:LinPolMap}). Indeed, while most of the reprocessed flux is still contained within the northern sky quadrant, we find that the polarisation degree is significantly higher in the east and west quadrants, so that the polarised flux is actually dominated by the contribution of these side quadrants, and the net polarisation is parallel to the torus axis ($\gamma=0^\circ$). At even higher energies, the polarised flux is dominated by photons escaping thought the obscuring front side of the torus, mostly from the east and west sky regions, so that $Q>0$ and $\gamma=0^\circ$.

The flux balance between the northern sky region ($Q<0$) and the east and west sky regions ($Q>0$) is determined by the level of obscuration of the latter regions. This explains why the transition energy $E_\text{flip}$ scales with the torus opacity (see Fig.~\ref{fig:spectra}), as higher torus column densities require higher photon energies (with more penetrating power) to escape from the east and west sky regions. This is a direct result of the specific torus geometry, and we conclude that spatially resolved Stokes surface brightness maps form a powerful tool to study the geometrical effects that are encoded in spectro-polarimetric observations. The SKIRT code allows for calculating these surface brightness maps at a high signal-to-noise, in limited computational time.

\subsection{Inclination - covering factor contours}
\label{sec:contours}
\begin{figure*}
    \centering
	\includegraphics[width=0.95\hsize]{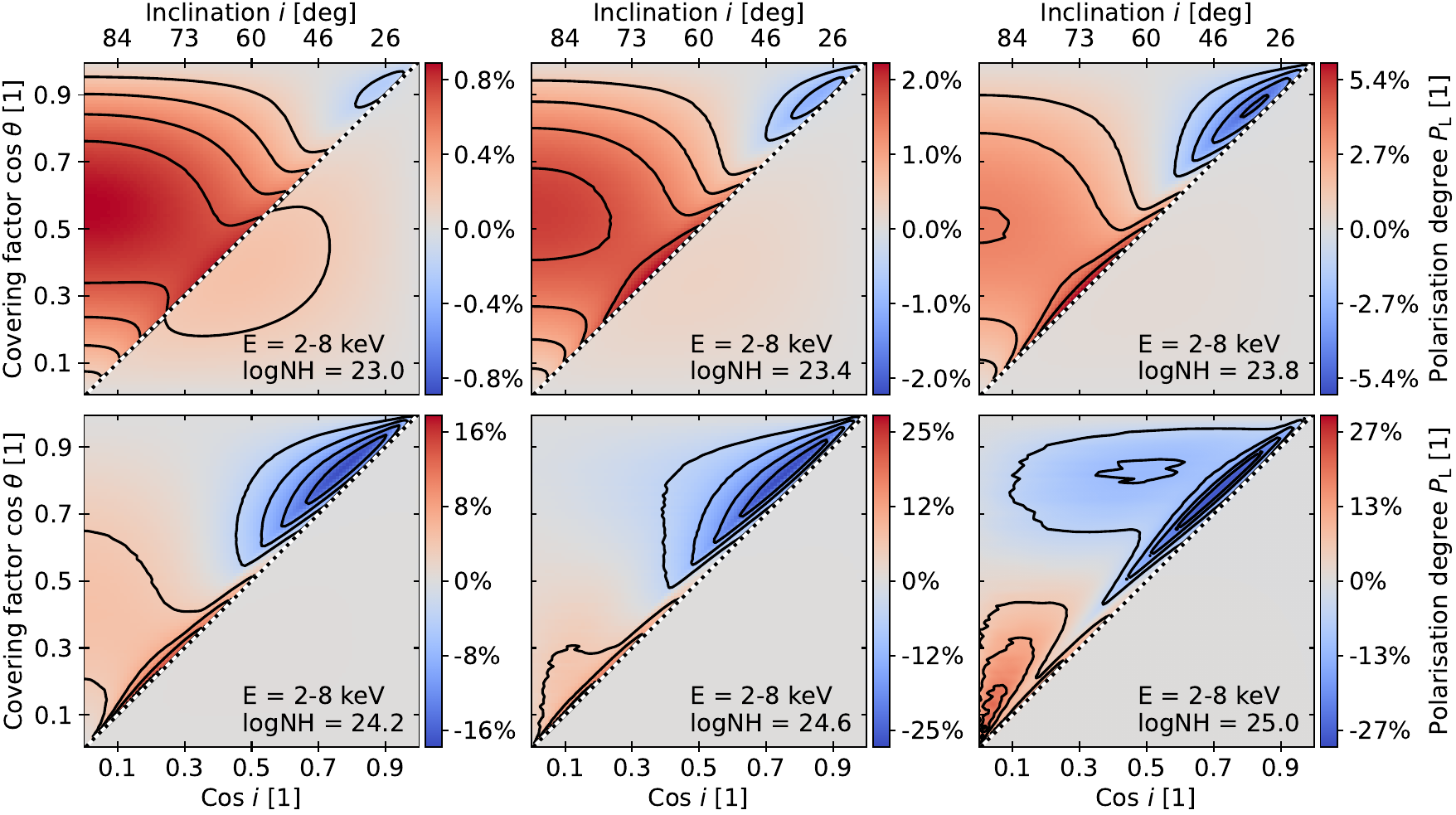}
    \caption{Linear polarisation degree $P_{\text{L}}$ in the $2-8~\text{keV}$ band for the wedge torus model described in Sect.~\ref{sec:setup} as a function of the torus covering factor $\text{CF}=\cos \theta$ and the observer inclination $\cos i$. The different panels represent different values for the torus column density $N_\text{H}$. Positive $P_{\text{L}}$ values (red) denote polarisation in the vertical direction of the observer frame ($\gamma=0^\circ$). Negative $P_{\text{L}}$ values (blue) denote polarisation in the horizontal direction of the observer frame ($\gamma=90^\circ$). The dotted line separates obscured and unobscured sightlines.}
    \label{fig:contours}
\end{figure*}
Inspired by Sect.~\ref{sec:polifoE}, where we found an interesting behaviour of the polarisation observables at $\cos i \lesssim \text{CF}$, this section discusses the wedge torus results of Sect.~\ref{sec:torus} as a function of the $\cos i$ and $\text{CF}$ model parameters. This analysis should also be useful for the interpretation of observational data, to constrain torus properties from broadband polarimetry. Fig.~\ref{fig:contours} shows the total polarisation degree (over the $2-8~\text{keV}$ band) as a function of $\cos i$ and $\text{CF}$, with different panels representing different values for the torus column density (increasing from $\log{N_\text{H}}=23$ to $\log{N_\text{H}}=25$).

We find that the total polarisation degree is virtually zero ($P_{\text{L}}<0.3\%$) for unobscured sightlines ($\cos i > \text{CF}$), where the polarisation signal of the AGN torus is mostly diluted by the direct source emission\footnote{This is when assuming unpolarised primary emission (Sect.~\ref{sec:setup}). Alternatively, the torus signal is `diluted' by the polarisation signal of the X-ray corona \citep[$P_{\text{L}}$ of a few percent; see e.g.][]{gianolli23}.}. The polarisation degree reaches a maximum at $\cos i \lesssim \text{CF}$, where the illuminated backside of the torus can be observed without significant obscuration, similar to Sect.~\ref{sec:polifoE}. Furthermore, the total polarisation degree is observed to increase from $1\%$ to $30\%$ when $\log{N_\text{H}}$ increases from $23$ to $25$, as more torus material results in more scattering interactions inducing a stronger polarisation signal, while also the unpolarised direct flux component is more obscured.

\begin{figure*}
    \centering
	\includegraphics[width=0.95\hsize]{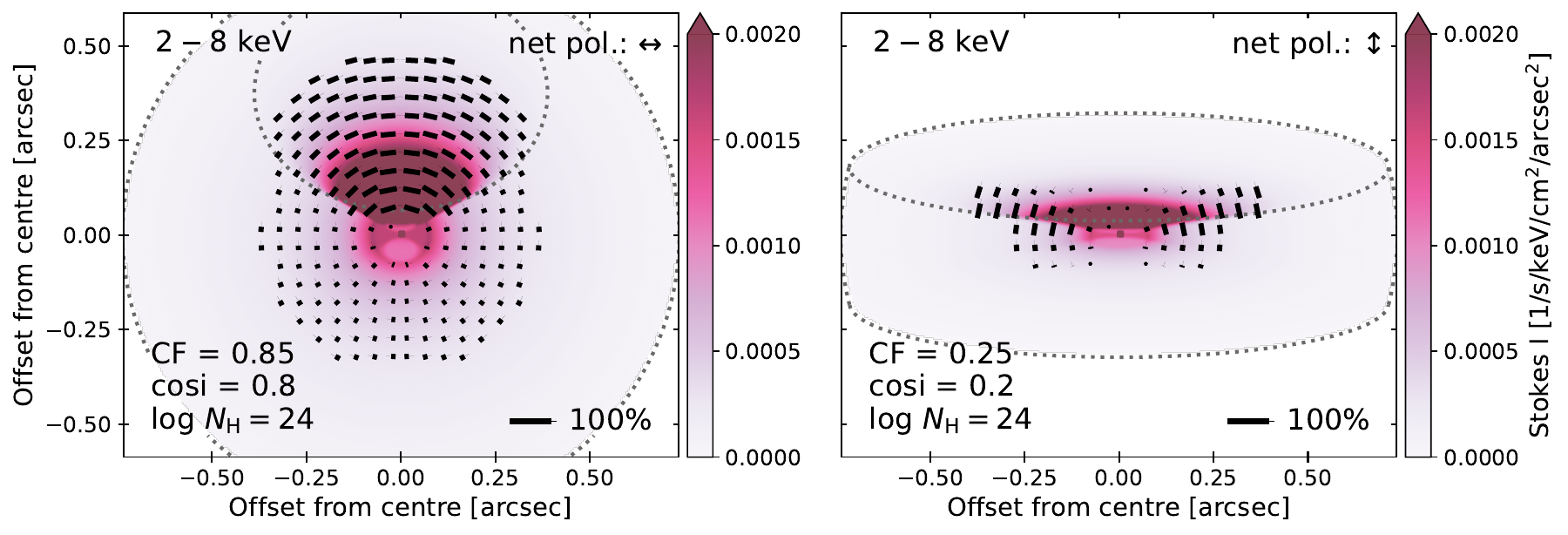}
    \caption{Linear polarisation maps for $\cos i \lesssim \text{CF} = 0.85$ (left) and $\cos i \lesssim \text{CF} = 0.25$ (right), visualising the spatially resolved polarisation degree and polarisation angle (similar to Fig.~\ref{fig:LinPolMap}, right) for the local polarisation maxima that were identified in Sect.~\ref{sec:contours} (see Fig.~\ref{fig:contours}). The background image represents the total surface brightness Stokes $I$ (in purple). The dotted grey lines indicate the torus outlines.
    }
    \label{fig:TwoRegions}
\end{figure*}
Finally, two distinct polarisation maxima can be observed in each panel of Fig.~\ref{fig:contours}, separated by a region where $P_{\text{L}}$ approaches zero. One local maximum is located at $\cos i \lesssim \text{CF} = 0.7 \textrm{ to } 0.9$ (top right corner of each panel, with $\gamma=90^\circ$), while the other local maximum is located at $\cos i \lesssim \text{CF} = 0.2 \textrm{ to } 0.4$ (bottom left corner of each panel, with $\gamma=0^\circ$). Similar to Sect.~\ref{sec:polifoE}, we calculated the spatially resolved Stokes surface maps to study this behaviour, which are shown in Fig.~\ref{fig:TwoRegions}. We find that both of these polarisation maxima are related to scattering on the illuminated backside of the torus, similar to Sect.~\ref{sec:polifoE}.

At $\cos i \lesssim \text{CF} = 0.7 \textrm{~to~} 0.9$ (i.e.\ tori that are almost entirely closed), the illuminated backside of the torus mostly covers the northern sky quadrant (where Stokes $Q<0$), so that the net polarisation signal is perpendicular to the torus axis ($\gamma=90^\circ$). For $\cos i \lesssim \text{CF} = 0.2 \textrm{ to } 0.4$ on the other hand (i.e.\ thin, disc-like tori), the illuminated backside mostly covers the east and west quadrants (where Stokes $Q>0$), so that the polarisation is parallel to the torus axis ($\gamma=0^\circ$). For parameter combinations in between these two maxima, the torus backside covers all three sky quadrant, so that positive and negative Stokes $Q$ contributions (partially) cancel out, eventually reaching $P_{\text{L}}=0$ at the border region that separates the two local maxima.

\subsection{Polarisation angle flip for different sightlines}
\label{sec:Eflip}
In Sect.~\ref{sec:results}, we presented Stokes spectra for one sightline only: $\cos i \lesssim \text{CF}$, which offers an unobscured view on the illuminated backside of the torus. The reason for this specific choice is that we plan to compare these results to simulations that include a polar component in follow-up work, showing similar spectra\footnote{Indeed, polar-extended dusty gas could act as an illuminated reflector that is visible through unobscured sightlines, which would require less fine-tuning of the observer inclination (forming a natural solution for strong reflection spectra).}. However, because of this very specific viewing angle, the trends that were found for $\text{CF}=0.45$ and $\cos i =0.4$ in Sect.~\ref{sec:results} might differ from more general trends at arbitrary obscured sightlines, which we investigate in this section. In particular, we focus on the photon energy $E_\text{flip}$ where the polarisation angle $\gamma$ flips from $90^\circ$ to $0^\circ$, as a function of the torus column density.

\begin{figure}
    \centering
	\includegraphics[width=\columnwidth]{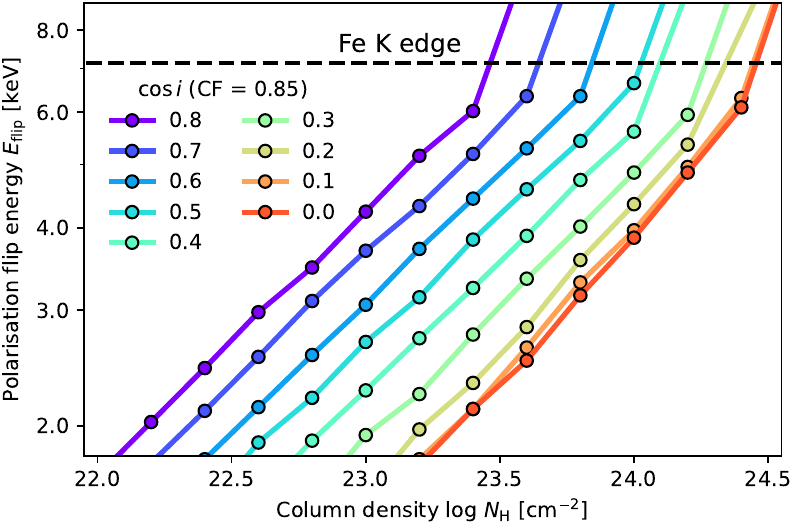}
    \caption{Photon energy $E_\text{flip}$ where the polarisation angle $\gamma$ flips from $90^\circ$ to $0^\circ$, as a function of the torus column density, for different observer inclinations. The source parameters are $\Gamma = 1.8$ and $E_\textrm{cut}=200~\text{keV}$ (same as Fig.~\ref{fig:spectra}), but the torus covering factor CF is now $0.85$. The Fe~K absorption edge at $7.1~\textrm{keV}$ is indicated in black.}
    \label{fig:Eflip_CF085}
\end{figure}
Using the radiative transfer results of the torus model grid that was presented in Sect.~\ref{sec:xspec}, we measured $E_\text{flip}$ for a broad range of torus parameter combinations, focussing on the same source parameters as Sect.~\ref{sec:results} ($\Gamma = 1.8$ and $E_\textrm{cut}=200~\text{keV}$), but a higher torus covering factor ($\textrm{CF}=0.85$) to allow for more obscured sightlines ($\cos i < \text{CF}$). In Fig.~\ref{fig:Eflip_CF085}, the polarisation flip energy $E_\text{flip}$ is shown as a function of the torus column density for different (obscured) observer inclinations, generalising the trend that was found in Sect.~\ref{sec:results}: A linear relation between $\log{N_\text{H}}$ and $\log{E_\text{flip}}$ is found for all viewing angles, which is broken at the Fe~K absorption edge at $7.1~\textrm{keV}$. Indeed, similar to Sect.~\ref{sec:results}, $E_\text{flip}$ is found to correlate with the torus opacity, which increases discontinuously at the Fe~K edge for a fixed column density. 

For $\textrm{CF}=0.85$, the polarisation flip happens when the torus medium becomes sufficiently transparent, so that the reprocessed flux escaping through the front side of the torus (mostly $Q>0$) starts to dominate over the flux related to scattering on the torus backside ($Q<0$). As the former region is obscured by the torus while the latter region is not, the ratio between these two components naturally scales with the torus opacity, explaining the trend with $\log{N_\text{H}}$ (see Fig.~\ref{fig:Eflip_CF085}). In addition, as the flux of the torus backside decreases with increasing inclination (which is a projection effect), the transmitted flux (having $Q>0$) starts to dominate at lower $E_\text{flip}$ for higher inclinations (i.e.\ lower $\cos i$), explaining the observed trend with $\cos i$ in Fig.~\ref{fig:Eflip_CF085}.

Finally, we note how the details of the polarisation angle flip depend on the specific combination of $\text{CF}$ and $\cos i$. For $\textrm{CF}=0.85$, we found a balance between transmission through the torus front side and scattering on the torus backside for all $\cos i$, which is different from the scenario discussed in Sect.~\ref{sec:polifoE}, where we found a balance between positive and negative Stokes $Q$ contributions on the torus backside for $\text{CF}=0.45$. For smaller covering factors, the positive Stokes $Q$ contributions on the torus backside could be dominating over the entire IXPE range (so that no polarisation flip is observed), while for some $\text{CF}$ and $\cos i$ combinations, two flips could occur. These intricacies, and their link to the torus geometry, can be studied in great detail with SKIRT, in particular based on Stokes surface brightness maps, which can be calculated in little computational time.

\subsection{XSPEC torus model}
\label{sec:discusxspec}
With this work, we release an AGN torus model that describes both X-ray spectra and X-ray polarisation for observational data fitting with XSPEC (see Sect.~\ref{sec:xspec}). This \textsc{xskirtor\textunderscore smooth} model represents a smooth toroidal reprocessor of cold gas, positioned in the equatorial plane, as described in Sect.~\ref{sec:setup} (see Fig.~\ref{fig:geometry}). While similar smooth torus models have been very successful in modelling the observational X-ray spectra of obscured AGN \citep[see e.g.][for recent examples]{kallova24, pizzetti24, layek24, zhao24, molina24}, these models do not incorporate geometrical complexities such as clumpy or filamentary substructures, or polar-extended dusty gas, which might be omnipresent in local AGN (Sect.~\ref{sec:intro}). Therefore, the \textsc{xskirtor\textunderscore smooth} model should be used as a tool to gain insights into the representative properties of the circumnuclear medium (such as its covering factor or average column density), more than the final solution for the AGN torus geometry, when being applied to observational data. 

The \textsc{xskirtor\textunderscore smooth} model is provided in two different flavour variations: a coupled configuration which provides the direct and reprocessed flux components as a single table, and a decoupled configuration which allows for varying the line-of-sight column density independently from the equatorial column density\footnote{In fact, all model parameters could be varied independently (given that there would be a physical motivation to do so).} \citep[see e.g.][]{torresalba23, pizzetti24}. As a final remark, we note that the exponential cut-off energy of the primary X-ray source (Sect.~\ref{sec:torus}) has a noticeable effect in the $2-8~\text{keV}$ range, even when $E_\textrm{cut}>100~\textrm{keV}$\footnote{For example, for $E_\textrm{cut} = 100~\text{keV}$, the effect of the exponential cutoff on the X-ray spectrum at $8~\text{keV}$ is 8\%.}. Therefore, the $\log{E_\textrm{cut}}$ model parameter (see Table~\ref{table:1}) is relevant beyond the modelling of hard X-ray spectra and should not be neglected at lower X-ray energies such as the IXPE range.

\section{Summary and outlook}
\label{sec:summary}
In this work, we presented a general framework for modelling X-ray polarisation in 3D radiative transfer simulations of cold gas and dust, to model the spectro-polarimetric signal that is produced by X-ray reprocessing in AGN circumnuclear media. We discussed how radiative transfer processes depend on the polarisation state of the incoming photon and how the polarisation state is updated by Compton scattering and fluorescence (Sect.~\ref{sec:framework}). We described how polarised X-ray radiative transfer can be implemented using a Monte Carlo method and provided an implementation to the 3D radiative transfer code SKIRT, which is publicly available online\footnote{\url{https://skirt.ugent.be}} (Sect.~\ref{sec:implementation}).

As a first application, we focussed on a 2D torus geometry in Sect.~\ref{sec:torus}, to demonstrate the new X-ray polarisation capabilities of the SKIRT code without going into the details of 3D structure and its effect on spectro-polarimetric observables. However, we note that the current SKIRT implementation works in full 3D already, so that more complex models \citep[such as those presented by][]{stalevski17, stalevski19, vandermeulen23} can be run with X-ray polarisation enabled (i.e.\ without any further modifications to the code). In future work, we will focus on these models with a truly 3D structure beyond the classical torus.

For the 2D wedge torus model (Sect.~\ref{sec:setup}), we calculated Stokes spectra at a high signal-to-noise (Fig.~\ref{fig:spectra}) and computed the linear polarisation angle and polarisation degree as a function of photon energy. We found that the polarisation angle flips from $90^\circ$ to $0^\circ$ at a specific energy inside the IXPE range (Sect.~\ref{sec:results}), which we interpreted as a balance between the reprocessed flux originating from different regions of the torus, with a direct link to the torus geometry (Sect.~\ref{sec:polifoE}). Furthermore, we found that the polarisation degree reaches a maximum at (obscured) sightlines having $\cos i \lesssim \text{CF}$ (Sect.~\ref{sec:contours}), where the torus backside can be observed without significant obscuration. However, depending on the torus covering factor, this polarisation maximum can be parallel or perpendicular to the torus axis (see Fig.~\ref{fig:TwoRegions}). Finally, we found that the specific photon energy $E_\text{flip}$ where the polarisation angle $\gamma$ flips from $90^\circ$ to $0^\circ$ scales with the torus opacity (Sect.~\ref{sec:Eflip}), as the polarisation flip is related to a balance between torus regions with different levels of obscuration. These intricacies, and their link to the torus geometry, were studied in great detail using the Stokes surface brightness maps, which SKIRT can calculate in a short amount of computational time.

With this work, we release spectro-polarimetric templates for fitting observational data of obscured AGN based on the torus model grid presented in Sect.~\ref{sec:xspec} (and discussed in Sect.~\ref{sec:discusxspec}). This  X-ray torus model is provided as an XSPEC table named \textsc{xskirtor\textunderscore smooth}, which can simultaneously describe X-ray spectra and spectro-polarimetry over the $0.3$ to $200~\text{keV}$ range, with a spectral resolution of $R=154$ (Sect.~\ref{sec:setup}). We provided an additional high-resolution XSPEC table model with an adaptive energy resolution over the $1.5$ to $15~\text{keV}$ subrange, for fitting the microcalorimeter X-ray spectra obtained with XRISM/Resolve and \emph{Athena}/X-IFU. All tables are publicly available online\footnote{\url{https://github.com/BertVdM/xskirtor}}.

The SKIRT code can now model X-ray polarisation in AGN circumnuclear media and predict the spectro-polarimetric X-ray signal of complex 3D models, with all features of the established SKIRT framework available. SKIRT is highly optimised in terms of computational efficiency, allowing for complex 3D models to be explored in a short timeframe. Furthermore, the SKIRT code offers an unmatched geometrical flexibility for setting up simulations in full 3D, which has now become available to X-ray polarisation modelling. Finally, SKIRT can calculate polarisation maps at a high signal-to-noise (see Fig.~\ref{fig:LinPolMap} and~\ref{fig:TwoRegions}), which forms a powerful tool to study the geometrical effects that are encoded in spectro-polarimetric observations. SKIRT can calculate fluxes, images, spectra and polarisation maps from mm to X-ray wavelengths, and the community is warmly invited to use the code in any way they see fit.

\begin{acknowledgements}
      B.\ V.\ acknowledges support by the Fund for Scientific Research Flanders (FWO-Vlaanderen, project 11H2121N). M.\ S.\ acknowledges support by the Science Fund of the Republic of Serbia, PROMIS 6060916, BOWIE and by the Ministry of Education, Science and Technological Development of the Republic of Serbia through the contract No. 451-03-9/2022-14/200002. G.\ M.\ acknowledges financial support from Italian MUR under grant PNRR-M4C2-I1.1-PRIN 2022-PE9-An X-ray view of compact objects in polarized light -F53D23001230006-Finanziato dall'U.E.-NextGenerationEU. We wish to thank K.\ A.\ Arnaud for support with the XSPEC package. 
\end{acknowledgements}
\bibliographystyle{aa}
\bibliography{finalrefs}

\end{document}